\documentclass[a4paper,11pt]{article}
\pdfoutput=1 

\usepackage{jheppub} 

\usepackage[T1]{fontenc} 
\usepackage{tikz-cd}
\usepackage{caption}
\usepackage{soul}
\usepackage{ulem}

\title{\boldmath  Classifying Causal Nonlinear Electrodynamics via $\varphi$-Parity and Irrelevant Deformations}

\author[a]{H. Babaei-Aghbolagh,}
\author[b]{Komeil Babaei Velni,}
\author[a,c]{Song He,}
\author[b]{Zahra Pezhman,}

\affiliation[a]{Institute of Fundamental Physics and Quantum Technology, \& School of Physical Science and Technology, Ningbo University, Ningbo, Zhejiang 315211, China}
\affiliation[b]{Department of Physics, University of Guilan, P.O. Box 41335-1914, Rasht, Iran}
\affiliation[c]{Max Planck Institute for Gravitational Physics (Albert Einstein Institute), Am M\"uhlenberg 1, 14476 Golm, Germany}
\emailAdd{hosseinbabaei@nbu.edu.cn}
\emailAdd{babaeivelni@guilan.ac.ir}
\emailAdd{ hesong@nbu.edu.cn}
\emailAdd{zpezhmanh@phd.guilan.ac.ir}

\abstract{We investigate the classification of self-dual nonlinear electrodynamic (NED) theories based on their analyticity properties, which are directly linked to invariance under a discrete $\varphi$-parity transformation. This classification is expressed through the structure of the irrelevant $T\bar{T}$-like deformations that generate the theories from a Maxwell seed. Using both closed-form and perturbative methods within the Courant-Hilbert (CH) and Russo-Townsend auxiliary field formalisms, we demonstrate a precise correspondence: $\varphi$-parity-invariant, analytic theories are generated by irrelevant deformations built from integer powers of the energy-momentum tensor scalars, $\mathcal{O}_\lambda \sim \sum C_m (T_{\mu\nu}T^{\mu\nu})^{1-m}({T_{\mu}}^{\mu}{T_{\nu}}^{\nu})^{m}$. Conversely, $\varphi$-parity-violating, non-analytic theories require deformations involving both integer and half-integer powers, $\mathcal{O}_\lambda \sim \sum C_m (T_{\mu\nu}T^{\mu\nu})^{1-m/2}({T_{\mu}}^{\mu}{T_{\nu}}^{\nu})^{m/2}$. We prove this result in generality via a perturbative CH framework, showing that $\varphi$-parity invariance imposes specific constraints on the expansion coefficients of the CH function $\ell(\tau)$ which, in turn, force all half-integer powers in the deformation to vanish. The classification is explicitly verified for known closed-form theories: the analytic generalized Born-Infeld model and the non-analytic examples of the $q=3/4$-deformed and "no $\tau$-maximum" theories. Furthermore, we show how the $\varphi$-parity transformation is consistently generalized in the presence of a marginal root-$T\bar{T}$ coupling $\gamma$, and we derive the corresponding marginal and irrelevant flow equations for the studied theories.
}

\begin{document}
\maketitle
\flushbottom
\section{Introduction}\label{0.}
Nonlinear electrodynamics (NED) was originally proposed by Born and Infeld in 1934 with the aim of constructing a theory in which the self-energy of point charges remains finite, and the divergence of the field near the charge is eliminated~\cite{Born:1934gh, Born:1933lls}. This nonlinear structure naturally emerges in string theory from the Dirac–Born–Infeld (DBI) action associated with D-branes~\cite{Fradkin:1985qd}. 
A critical consideration in many electrodynamic applications is causality (Waves should not propagate faster than light), which plays a fundamental role in ensuring physical consistency. 
Recent developments have shown that NED provides a consistent setting in which key physical principles, such as the causal propagation of fields and compatibility, can be realized within a single unified framework~\cite{Russo:2024llm,Russo:2024xnh, Russo:2024ptw}. Within this context, one can construct Lagrangian densities that remain causal and preserve the fundamental symmetries of Maxwell theory, including gauge invariance, and electric-magnetic duality~\cite{ Bialynicki-Birula:1992rcm,Gaillard:1981rj, Gaillard:1997rt,Gibbons:1995ap,Gibbons:1995cv,Avetisyan:2021heg}.

The self-duality condition for nonlinear electrodynamics, when expressed in its differential form, constitutes a fundamental constraint on the Lagrangian density $\mathcal{L}$. This constraint takes the form of a nonlinear partial differential equation (PDE) governing the functional dependence of $\mathcal{L}$ on the two fundamental Lorentz-invariant scalar variables. The specific form of this PDE is given by~\cite{Bialynicki-Birula:1992rcm,Gaillard:1981rj,Gaillard:1997rt,Gibbons:1995ap,Gibbons:1995cv,Kuzenko:2026kvr,Avetisyan:2021heg}:
\begin{equation}
\label{eq:lagrange_self_duality}
\left( \partial_S \mathcal{L} \right)^2 - 2 \frac{S}{P} \left( \partial_S \mathcal{L} \right) \left( \partial_P \mathcal{L} \right) - \left( \partial_P \mathcal{L} \right)^2 = 1\,,
\end{equation}
where the derivatives $\partial_S \mathcal{L}$ and $\partial_P \mathcal{L}$ denote the partial derivatives of the Lagrangian with respect to the electromagnetic invariant variables $S$ and $P$. These invariants are defined as:
\begin{align}
S &= -\frac{1}{4} F_{\mu\nu} F^{\mu\nu} = \frac{1}{2} (\vec{E}^2 - \vec{B}^2), \\
P &= -\frac{1}{4} F_{\mu\nu} \widetilde{F}^{\mu\nu} = \vec{E} \cdot \vec{B},
\end{align}
which are constructed from the field strength tensor $F^{\mu\nu}$ and its Hodge dual $\widetilde{F}^{\mu\nu} = \frac{1}{2} \epsilon^{\mu\nu\rho\sigma} F_{\rho\sigma}$. A Lagrangian $\mathcal{L}(S, P)$ satisfying Equation~\eqref{eq:lagrange_self_duality} describes a theory whose equations of motion are invariant under the duality transformation rotations in the space of electric and magnetic fields, a property characteristic of self-dual theories such as the well-known Born-Infeld electrodynamics. 
A reformulation of the duality-invariant condition in~\eqref{eq:lagrange_self_duality} can be obtained via the following change of variable:
\begin{equation}\label{UV}
	U=\frac{1}{2}(\sqrt{S^2+P^2}-S)
	,\quad
	V=\frac{1}{2}(\sqrt{S^2+P^2}+S),
\end{equation}
where U and V are non-negative scalars.
The condition that $\mathcal L$ must satisfy to exhibit self-duality assumes the following simple form when $\mathcal L$ is expressed as a function of  
(U, V) \cite{Gaillard:1997rt,Gibbons:1995cv}:
\begin{eqnarray}\label{sdc}
    {\cal L}_U {\cal L}_V=-1.
\end{eqnarray}
Several approaches exist for solving the differential equation~\eqref{sdc} and identifying duality-invariant electrodynamic theories. Well-known methods include the Courant-Hilbert approach~\cite{10.1115/1.3630089}, the auxiliary field method~\cite{Ivanov:2002ab,Ivanov:2003uj}, the framework presented in~\cite{Mkrtchyan:2022ulc}, the perturbation technique~\cite{Babaei-Aghbolagh:2024uqp,Babaei-Aghbolagh:2025cni}, as well as the new auxiliary field approach recently introduced by Russo and Townsend~\cite{Russo:2025fuc}. This work compares the Courant-Hilbert and the new auxiliary field approaches. We explain the causality condition of the theory from the Courant-Hilbert approach and demonstrate how the $\varphi$-parity invariance within the new auxiliary field approach leads to the generation of analytic theories\cite{Russo:2024llm,Russo:2024ptw,Russo:2024xnh,Russo:2025fuc}.

Inspired by solvable irrelevant deformations in two-dimensional quantum field theories~\cite{Smirnov:2016lqw,Cavaglia:2016oda}, known as $T\bar{T}$ deformations, this mechanism has recently been generalized to higher dimensions and to theories with gauge fields~\cite{Conti:2018jho,Babaei-Aghbolagh:2020kjg,Babaei-Aghbolagh:2022uij,Babaei-Aghbolagh:2022itg,Ferko:2023wyi,Ferko:2024zth,Hutomo:2025dfx}. This formulation of deformations is driven by a series function of the two structures, the energy-momentum tensor~\cite{Babaei-Aghbolagh:2024uqp}. When applied to the Maxwell Lagrangian, it generates causal, duality-preserving theories in closed form, such as Born-Infeld theory.
We will demonstrate that the introduction of $T\bar{T}$ deformations into causal electromagnetic theories generically leads to two distinct classes of irrelevant deformations, corresponding to the analytic and non-analytic theories. The first class, associated with the analytic theories, is characterized by deformations involving only integer powers of momentum-energy structures. In contrast, the second class, which describes non-analytic theories, is distinguished by the inclusion of both integer and half-integer powers.

The deformed theory is characterized by an irrelevant operator built from two fundamental momentum-energy structures: the $\left(T_{\mu\nu}T^{\mu\nu}\right)$ and the $\left({T_{\mu}}^{\mu} {T_{\nu}}^{\nu}\right)$ operator. As a fundamental result, we have demonstrated that:
\begin{itemize}
    \item For \textbf{analytic theories}, the irrelevant $T\bar{T}$ deformations are spanned by \textbf{integer powers}:
  \begin{eqnarray}\label{AN}
    \mathcal{O}^{\text{(analytic)}}_{\lambda^{2m}} \sim  \sum_{m=0}^{\infty} C_m \left(T_{\mu\nu}T^{\mu\nu}\right)^{1-m} \left({T_{\mu}}^{\mu} {T_{\nu}}^{\nu}\right)^{m}, \qquad \text{for } m \in \mathbb{Z}_{\geq 0}.
\end{eqnarray}
    \item For \textbf{non-analytic theories}, the irrelevant $T\bar{T}$ deformations are spanned by \textbf{both integer and half-integer powers}:
   \begin{eqnarray}\label{nonAN}
    \mathcal{O}^{\text{(non-analytic)}}_{\lambda^{m}} \sim  \sum_{m=0}^{\infty} C_m \left(T_{\mu\nu}T^{\mu\nu}\right)^{1-\tfrac{m}{2}} \left({T_{\mu}}^{\mu} {T_{\nu}}^{\nu}\right)^{\tfrac{m}{2}}, \qquad \text{for } m \in \mathbb{Z}_{\geq 0}.
   \end{eqnarray}
\end{itemize}
This result provides an explicit, general conclusion for all known theories studied using the Courant-Hilbert approach and the new auxiliary-field formalism, thereby clarifying the connection between irrelevant $T\bar{T}$ deformations and analytic theories. In this work, we explore theories beyond Born-Infeld and logarithmic electrodynamics, including $q$-deformed and no $\tau$-maximum theories. We develop a general framework for these extensions and derive their irrelevant and marginal flow equations using perturbative techniques.

The organization of this paper is as follows. In Section~\eqref{02}, we review the foundational formalisms for self-dual nonlinear electrodynamics, presenting both the Courant-Hilbert approach, which connects the Lagrangian to an arbitrary function $\ell(\tau)$, and the newer auxiliary field method of Russo and Townsend. We derive the causality conditions and introduce the energy-momentum tensor structures central to $T\bar{T}$ deformations. We introduce the framework of irrelevant $T\bar{T}$-like deformations, including the marginal root-$T\bar{T}$ flow. In Section~\eqref{033}, we analyze closed-form models such as generalized Born-Infeld, $q=3/4$-deformed, and ``no $\tau$-maximum'' theories, establishing their $\varphi$-parity properties and explicitly deriving their irrelevant deformation operators. Section~\eqref{0044} presents a general perturbative framework, analyzing $q$-deformed theories and constructing a master ``Courant-Hilbert perturbation theory.'' We prove that $\varphi$-parity invariance imposes specific coefficient constraints that eliminate half-integer powers from the deformation, thus characterizing all analytic theories. We conclude in Section~\eqref{0055} with a summary of results and an outlook on future research directions.

\section{Self-dual nonlinear electrodynamics formalism}\label{02}

\subsection{Courant-Hilbert approach  and causality }\label{021}
The general solution to  \eqref{sdc} is  implicitly expressed in terms of the variables (U, V) and an arbitrary  function of real value $\ell(\tau)$ of a real variable~\cite{10.1115/1.3630089}:
\begin{eqnarray}\label{Landtau}
	\mathcal{L}=\ell(\tau)-\frac{2U}{[\dot{\ell}(\tau)]},\quad
	\tau=V+\frac{U}{[\dot{\ell}(\tau)]^2},
\end{eqnarray}
where $\dot{\ell}=\frac{d\ell}{d\tau}$. Note that $\tau \geq 0$ by definition, with equality only for $U = V = 0$. 
The choice $\ell(\tau) = \tau$ yields the free-field Maxwell case, 
with ${\cal L}_U =-1/\dot{\ell}(\tau)$ and $ {\cal L}_V=\dot{\ell}(\tau)$ satisfying the self-duality condition ~\eqref{sdc}.

Any NED theory that is causal in the weak-field regime but fails to remain causal for sufficiently strong fields uniquely determines a new causal and self-dual NED theory that shares the same Lagrangian density in the absence of a magnetic field. 
\subsection*{causality condition:}
In self-dual theories, causality in the strong-field regime is a consequence of causality in the weak-field regime\cite{Russo:2024llm}. A self-dual NED with a weak-field limit is causal if and only if \cite{Russo:2024llm}:
\begin{eqnarray}\label{cc}
   \dot\ell\ge1,\quad
\ddot\ell\ge0
\end{eqnarray}
where  $\ddot\ell\ge0$ implies that $\mathcal{L}(S, P)$ is a convex function of $S$ and $P$. For  general NED, assuming zero vacuum energy, causality
implies both the Dominant Energy Condition (DEC) and the Strong
Energy Condition (SEC). For self-dual NED theories, weak-field
causality alone implies both the DEC and the SEC \cite{Russo:2024xnh}.

The energy-momentum tensor (EMT) corresponding to the Lagrangian given in equation \eqref{Landtau} is expressed as \cite{Russo:2024xnh, Russo:2024ptw}:
\begin{eqnarray}\label{enerr}
	T_{\mu\nu} = \left(\frac{\tau\dot\ell}{U+V}\right) T^{\rm Max}_{\mu\nu} +
	(\ell - \tau \dot\ell) {\rm g}_{\mu\nu} \, ,  
\end{eqnarray}
where $T^{Max}_{\mu\nu}=F_{\mu\rho}{F_{\nu}}^{\rho}-\frac{1}{4} \,g_{\mu\nu} F_{\alpha\beta}F^{\alpha\beta}$ denotes the EMT of Maxwell's electrodynamics.
 As a result, the trace of the EMT in a general self-dual electrodynamic theory can be expressed as follows~\cite{Russo:2024xnh}:
\begin{eqnarray}\label{TnNl}
	{T_{\mu}}^\mu = 4(\ell - \tau \dot\ell) .
\end{eqnarray}
The condition $ {T_{\mu}}^\mu  = 0 $ holds if and only if $ \ell = \tau \dot\ell $. From the energy-momentum tensor given in Eq.~\eqref{enerr}, we can derive two distinct structures:
\begin{eqnarray} \label{ttg}  
	T_{\mu\nu}T^{\mu\nu}  =  4\big[(\tau\dot\ell)^2 + (\ell -\tau\dot\ell)^2\big], \qquad {T_{\mu}}^{\mu} {T_{\nu}}^{\nu}= 16(\ell - \tau\dot\ell)^2.   
\end{eqnarray}
Since all Lorentz invariants constructed from the stress tensor can be expressed as functions of ${T_{\mu}}^\mu $ and $T_{\mu\nu}T^{\mu\nu} $, they can also be written as functions of $\tau$. 
In four dimensions, the electromagnetic field strength possesses only two independent Lorentz invariants, $S$ and $P$. Consequently, any nonlinear electrodynamics Lagrangian and its associated energy--momentum tensor depend exclusively on these two scalars. This restricted dependence implies that the eigenvalue structure of $T_{\mu\nu}$ is degenerate, with only two independent eigenvalues. As a result, higher-order invariants such as $\mathrm{tr}(T^3)$, $\mathrm{tr}(T^4)$, and $\det T$ are not independent, but are algebraically determined by the quadratic combinations $T_{\mu\nu}T^{\mu\nu}$ and ${T_{\mu}}^{\mu} {T_{\nu}}^{\nu}$.

This structural simplification explains why all consistent irrelevant deformations of nonlinear electrodynamics, including $T\bar{T}$-like, root-$T\bar{T}$, and related flows, can be formulated solely in terms of these two invariants. The result is not accidental, but rather a direct consequence of Lorentz symmetry, the algebraic properties of electromagnetic fields in four dimensions, and the constrained form of the stress--energy tensor. This property enables the construction of $T\bar{T}$-like deformation operators, which we will discuss in the section~\eqref{0222}.
\subsection{New auxiliary field formalism and $\varphi$-parity}
A novel formulation of self-dual nonlinear electrodynamics is presented in Refs.~\cite{Russo:2024ptw, Russo:2025fuc} as an alternative to the CH constructions described in Section~\eqref{021}. The central element of this formulation is an auxiliary pseudo-scalar field $\varphi$, whose interactions are governed by a chosen potential function $W(\varphi)$. This framework is designed to satisfy the self-duality condition. Furthermore, it is demonstrated that imposing the causality principle on the potential $W(\varphi)$ is sufficient to guarantee the existence of a unique and physically consistent solution to the equation of motion for the auxiliary field. 
In this approach, the Lagrangian density takes the form
\begin{equation}\label{lsp}
    \mathcal{L}(S, P; \varphi) 
    = \cosh\varphi\, S 
    + \sinh\varphi\, \sqrt{S^2 + P^2} 
    - W(\varphi).
\end{equation}
 This formulation provides a unified and analytic description of self-dual NED.
The self-duality condition is well known in the classical literature and bears resemblance to the integrability condition~\cite{Babaei-Aghbolagh:2025hlm,Babaei-Aghbolagh:2025uoz}. 
A completely new approach, based on the auxiliary field, has been introduced in Ref.~\cite{Russo:2025fuc} to satisfy equation~\eqref{sdc}. This method develops a simplified formulation of self-dual nonlinear electrodynamics (NED) by introducing a new auxiliary pseudoscalar field, $\varphi$, and a potential function, $W(\varphi)$.
Considering the two variables $ U$ and $ V$, the Lagrangian~\eqref{lsp} can be rewritten as:
\begin{equation}\label{parityuv}
\mathcal{L}(U,V,y)
= -\frac{U}{y} \;+\; y\,V \;-\; \Omega(y), \qquad
\Omega'(y)=\frac{U}{y^{2}}+V,
\end{equation}
with the definitions $y = e^\varphi$ for auxiliary field $\varphi$ and $\Omega(y)$ as the master potential, the latter being fixed by the dualization constraints.
A one-to-one correspondence is established between the Courant-Hilbert approach and the new auxiliary field method as follows:
\begin{equation}
\ell(\tau) = \tau y - \Omega(y), \quad \tau = \Omega'(y), \quad y = \ell'(\tau),
\label{eq:legendre}
\end{equation}
Using this correspondence, we can express the two momentum-energy structures of~\eqref{ttg} as follows:
\begin{equation}
	\label{Tnunu1}
	T_{\mu\nu}T^{\mu\nu}=4 \Bigl(\Omega^2 (y) + y^2 {\Omega^{\prime}}^2 (y)\Bigr)\,\,, \,\,\,\,\,\,\,\,\,\,\,\, {T_{\mu}}^{ \mu}\,{T_{\nu}}^{ \nu}=16  \Omega ^2(y).
\end{equation}
\subsubsection*{Analytic  theories and $\varphi$-parity}
The analyticity of the weak-field expansion is equivalent to the invariance of the potential under the $\varphi$-parity transformation~\cite{Russo:2024ptw}:
$
    \varphi \rightarrow -\varphi.
$ These correspond precisely to models with even potentials, satisfying
\begin{equation}\label{exaparit}
    W(\varphi) = W(-\varphi).
\end{equation}
In their new formulation of self-dual electrodynamics, Russo and Townsend demonstrate that the imposition of $\varphi$-parity, restricting the auxiliary field potential to be an even function, is the crucial condition that ensures the theory yields a Lagrangian description for a generic {\it analytic} self-dual nonlinear electrodynamics theory.
As a result, the $\varphi$-parity transformations for any duality invariant electromagnetic theory take the form of $y \to 1/y$.
By definition, for any  analytic theory, the function $\Omega$ must satisfy the following symmetry~\cite{Russo:2025fuc}:
\begin{equation}\label{y1/y}
\Omega(y) =\hat{ \Omega}(y^{-1}),
\end{equation}
which reflects an intrinsic symmetry under inversion. This property leads directly to the transformation of the Lagrangian:
\begin{equation}\label{uvuv}
\mathcal{L}(U, V) = \mathcal{L}(-V, -U).
\end{equation}
Such a relation is consistent with expectations, as it corresponds to the effect of a $\varphi$-parity transformation applied after the elimination of the auxiliary field. The interchange of $U$ and $V$ with their negatives encapsulates the $\varphi$-parity-induced reversal of field components, preserving the  $\varphi$-parity invariant structure of the theory.

The study of Modified Maxwell (ModMax) theory~\cite{Bandos:2020jsw} clarifies the relationship between $\gamma$ coupling and $\varphi$-parity couplings, demonstrating how a nonlinear electrodynamic theory with an unrelated coupling can be extended to one with both $\lambda$ and $\gamma$ couplings. In the CH approach, Maxwell's theory is derived as a seed theory from the CH function $\ell(\tau)= \tau$. Extending Maxwell's theory to ModMax theory is achieved by rescaling this CH function by $e^\gamma $, which takes the form $\ell(\tau)=e^\gamma  \tau$. This yields the following Lagrangian density:
\begin{eqnarray}
    {\mathcal L}_{MM}=e^\gamma V-e^{-\gamma} U\,.
\end{eqnarray}
As shown in Ref.~\cite{Babaei-Aghbolagh:2022uij}, ModMax theory obeys the marginal flow equation $\partial_\gamma \,{\mathcal L}_{MM}
= \, \mathcal{R}_\gamma$. 
This remains true for all theories with ModMax as the weak-field limit because $\varphi$-parity flips the sign of $\gamma$. We note that $\varphi$-parity has again flipped the sign of $\gamma$; for weak fields, the $\varphi$-parity dual theory becomes ModMax but with $\gamma \to -\gamma$.
 A general formalism for extending a theory with a $\lambda$ coupling to one with two couplings $(\lambda,\gamma)$ can be found in Ref.~\cite{Babaei-Aghbolagh:2025uoz}. With the addition of $\gamma$ coupling, the flow equations in ~\eqref{marg} become valid for any theory, and the marginal flow equation can be determined explicitly.
The rescaling associated with $\Omega(y)$ is  by the ratio $\varphi \to \tilde{\varphi}  =\varphi-\gamma$. 
This definition necessitates a corresponding redefinition of the $\varphi$-parity  in the presence of the $\gamma$ coupling: $ \tilde{\varphi}\to - \tilde{\varphi}$. 
Therefore, the $\varphi$-parity transformations invariant in Eqs.~\eqref{y1/y} and \eqref{uvuv} for analytic theories are redefined as follows in the presence of the $\gamma$ coupling:
\begin{eqnarray}\label{pariii}
\Omega(y,\gamma) =\hat{ \Omega}(y^{-1},  -\gamma) \quad  \Longrightarrow  \quad\mathcal{L}(U, V,  \gamma) =\mathcal{L}(-V, -U,-  \gamma).
\end{eqnarray}
The transformations given in \eqref{pariii} represent the generalized form of $\varphi$-parity in the presence of $\gamma$ pairing.
Therefore, there is a subtle distinction between theories that preserve \(\phi\)-parity Eq.~\eqref{pariii} in the presence of $(\gamma)$ coupling and truly analytic theories. Specifically, a theory that preserves \(\varphi\)-parity is analytic if and only if \((\gamma = 0)\). With this understanding, we will continue to refer to \(\varphi\)-parity in~\eqref{pariii} preserving theories as analytic.
\subsection{ $T\bar{T}$-like deformations}\label{0222}
References~\cite{Smirnov:2016lqw,Cavaglia:2016oda} have shown that a broad class of interesting physical observables can be computed explicitly from such a non-local UV theory. By introducing the $T\bar{T}$ operator, an irrelevant deformation constructed from the product of the holomorphic and antiholomorphic components of the energy-momentum tensor, $T \equiv T_{zz}$ and $\bar{T} \equiv T_{\bar z \bar z}$, one obtains a Lorentz-invariant deformation of the action,
\begin{equation}
S_{\mathrm{QFT}} = S_{\mathrm{CFT}} + S_{\lambda},
\qquad
S_{\lambda} = \lambda \int d^{2}x \, T\bar{T}.
\end{equation}
Within this framework, the deformed theory satisfies the flow equation
$\partial_\lambda S_{\mathrm{QFT}}
= \int d^{2}x \, (T\bar{T})_{\lambda},
$ which holds nonperturbatively for finite values of the coupling~$\lambda$.

This construction admits a natural extension to four-dimensional gauge theories, most notably Maxwell electrodynamics. In four dimensions, two closely related extensions arise: the irrelevant $T^2$ deformation~\cite{Conti:2018jho,Babaei-Aghbolagh:2020kjg,Babaei-Aghbolagh:2022uij,Babaei-Aghbolagh:2022itg,Ferko:2023wyi,Ferko:2022iru} and the root-$T\bar{T}$ deformation~\cite{Babaei-Aghbolagh:2022uij}. For a U(1) gauge theory with field strength $F_{\mu\nu}$ and energy-momentum tensor
$
T_{\mu\nu}
= F_{\mu\rho} F_{\nu}{}^{\rho}
- \frac{1}{4} g_{\mu\nu} F_{\rho\sigma} F^{\rho\sigma},
$
the irrelevant $T^2$ operator is defined as
\begin{equation}
\mathcal{O}_\lambda
=\frac{1}{8} \big( T_{\mu\nu} T^{\mu\nu}
- \frac{1}{2} T^\mu{}_\mu T^\nu{}_\nu\big)
\end{equation}
and the flow is generated by
\begin{equation}\label{feqmax}
\partial_\lambda S_\lambda
= \int d^{4}x \, \mathcal{O}_\lambda \qquad
S_{\delta\lambda }
= S_{\text{Maxwell}}
+\delta \lambda \int d^{4}x \, \mathcal{O}_\lambda \,.
\end{equation}
The solution of the flow equation in ~\eqref{feqmax} yields Born-Infeld theory with a coupling $\lambda$. 
More remarkably, one may define the root-$T\bar{T}$ operator in four dimensions as~\cite{Babaei-Aghbolagh:2022uij}:
\begin{equation}\label{roootrr}
\mathcal{R}_\gamma
= \frac{1}{2} \sqrt{
T_{\mu\nu} T^{\mu\nu}
- \frac{1}{4} T^\mu{}_\mu T^\nu{}_\nu
}
\end{equation}
which generates a one-parameter family of deformed theories, with dimensionless $\gamma$-coupling,  through the marginal flow equation~\cite{Babaei-Aghbolagh:2022uij}:
\begin{equation}\label{marg}
\partial_{\gamma }S_\gamma
= \int d^{4}x \, \mathcal{R}_\gamma,
\qquad
S_{\delta\gamma }
= S_{\text{Maxwell}}
+ \delta\gamma \int d^{4}x \, \mathcal{R}_\gamma \,.
\end{equation}
The resulting root-$T\bar{T}$ deformation leads to the exactly solvable  ModMax theory, a nonlinear, conformal, and electromagnetic-duality-invariant extension of Maxwell electrodynamics. In contrast to the $T^2$ deformation, which introduces quartic interactions in field strength, the root-$T\bar{T}$ deformation produces a linear term in $\lvert F^2 \rvert$. 

Perturbative generalizations of irrelevant flow equations beyond the $T^2$ operator have been explored for other perturbative theories in two and four dimensions in Refs.~\cite{Babaei-Aghbolagh:2025lko,Babaei-Aghbolagh:2024uqp}. In particular, logarithmic electrodynamic theory, including $(\lambda,\gamma)$ couplings, obtained via the Courant-Hilbert formalism with the Courant-Hilbert function $	\ell (\tau)=- \frac{1}{\lambda} \log(1 -  e^{\gamma} \lambda \tau )$, has been studied in perturbation form in Ref.~\cite{Babaei-Aghbolagh:2025cni}, where its associated flow equation was derived.

A central objective of this work is to classify theories under $\varphi$-parity \eqref{pariii} as analytic or non-analytic by studying their irrelevant $T\overline{T}$ deformations. We demonstrate that the irrelevant operator for analytic theories adheres to the structure in \eqref{AN}, characterized by integer powers of momentum-energy terms. Conversely, for non-analytic theories, the operator takes the form in \eqref{nonAN}, which involves half-integer powers of these structures. We will establish the general form of these flow equations. We will classify the irrelevant flow equations into two categories based on $\varphi$-parity: those describing analytic theories and those describing non-analytic theories.

\section{Closed form models: irrelevant deformations and $\varphi$-parity}\label{033}
There are a few theories for which a closed-form solution has been derived using the CH formalism or the new auxiliary field approach. The scarcity of such theories stems from the challenge of solving the equations of motion with respect to the auxiliary field, a step which requires solving the second equation ~\eqref{parityuv}.

In this section, we examine several of these closed-form theories, with two couplings $(\lambda,\gamma)$, under $\varphi$-parity and identify their corresponding irrelevant deformations. We demonstrate that theories which are invariant under $\varphi$-parity, and thus considered analytic, possess irrelevant deformations consisting of integer powers of the $\left(T_{\mu\nu}T^{\mu\nu}\right)$ and  $\left({T_{\mu}}^{\mu} {T_{\nu}}^{\nu}\right)$ composite operators built from the energy-momentum tensor.
\subsection{GBI theory }
We now turn to theories that possess a closed form. The canonical example, and the best-known theory featuring two couplings $(\gamma, \lambda)$, is the general Born-Infeld theory. This theory is described by the CH function $\ell (\tau) = \tfrac{1}{\lambda}- \sqrt{\frac{1}{\lambda}\left(\tfrac{1}{\lambda}-2e^{\gamma}\tau\right)}$ within the Courant-Hilbert framework, and equivalently by the following potential in the new auxiliary field formulation:
\begin{eqnarray}\label{BIYY}  
W(\varphi) = \frac{1}{\lambda} \big( \cosh(\varphi-\gamma)-1\big) \quad \xrightarrow{\rm y= e^{\varphi}} \quad  \Omega(y)= \frac{1}{2\lambda} \Big((\frac{e^{\gamma}}{ y } + e^{-\gamma}y)-2 \Big).
\end{eqnarray}
It is evident that the $\Omega(y)$ potential in \eqref{BIYY} is invariant under the $\varphi$-parity transformations defined in \eqref{pariii}. Therefore, the general Born-Infeld theory is well known as an analytic theory. Moreover, the general Born-Infeld Lagrangian and its associated flow equations are well established. As shown in \cite{Conti:2018jho,Babaei-Aghbolagh:2020kjg,Babaei-Aghbolagh:2025cni}, the irrelevant flow equations of this theory are constructed from integer powers of the energy-momentum tensor and constitute an explicit example of the operators in \eqref{AN} with coefficients $C_m = 0, \quad m\geq2$.
 \subsection{q-deformed theory for $q=3/4$}
The $q$-deformed theory constitutes a one-parameter family of nonlinear electrodynamics (NED) models, with integrable cases generally constructed from the CH function:
\begin{eqnarray}  \label{qdeflll}
\ell(\tau)=\frac{1}{\lambda}-\frac{1}{\lambda}(1-\frac{1}{q}e^{\gamma} \lambda\tau)^q.
\end{eqnarray}
This theory exhibits two key limits: when $q = 1$, the CH function reduces to $ \ell(\tau)= e^{\gamma} \tau$, and likewise when $\lambda=0$, it also yields $\ell(\tau)= e^{\gamma} \tau$. In both limits, the $q$-deformed theory reduces to the ModMax theory.

Determining the closed form of q-deformed theories depends on solving the second equation \eqref{Landtau} and finding the closed form of $\tau$, which is not possible for every q-deformed theory.  
For  specific case $q = 3/4$, the CH function is given by $\ell(\tau)=\frac{1}{\lambda}-\frac{1}{\lambda}(1-\frac{4}{3}e^{\gamma}\lambda\tau)^\frac{3}{4}.$, and the second equation \eqref{Landtau} becomes solvable. Solving this yields the following closed form for $\tau$: 
 \begin{eqnarray}\label{tau2}
    \tau=\frac{1}{3}e^{-3\gamma}U(3e^\gamma\Lambda-2\lambda U)+V, 
\end{eqnarray}
where
$ \Lambda=\sqrt{{1-\lambda\,\frac{4 V e^{\gamma}}{3}+\lambda^2\,\Big(\frac{2e^{-\gamma}U}{3}\Big)^2}}.
$
By inserting the solution for $\tau$ from \eqref{tau2} into the Lagrangian\eqref{Landtau}, the resulting form of the $\frac{3}{4}\text{-deform}$ theory is:
 \begin{eqnarray}\label{L2}
	{\cal L}_{q=3/4}=\frac{1}{\lambda}-\frac{1}{\lambda}\sqrt{\Lambda-\frac{2e^{-\gamma}\lambda U}{3}}(\Lambda+\frac{4 e^{-\gamma}\lambda U}{3}).
   \end{eqnarray}
   This Lagrangian obeys the self-duality equation \eqref{sdc} and satisfies the causality condition \eqref{cc}. 
   By expressing the Lagrangian in closed form in terms of the two variables $ U $ and $ V $ in \eqref{L2}, we can derive consistent marginal and irrelevant $T\bar{T}$-like flow equations. First, we begin by examining the root-$ T\bar{T} $ flow equation.
   \subsection*{Root flow equation:}
In references~\cite{Babaei-Aghbolagh:2025cni, Babaei-Aghbolagh:2025uoz}, which detail the general theory derived via the CH approach from the Lagrangian in~\eqref{Landtau}, the  root operator~\eqref{roootrr} can be expressed as
\begin{eqnarray}
  	\mathcal{R}_\gamma
=\frac{1}{2} \sqrt{
T_{\mu\nu} T^{\mu\nu}
- \frac{1}{4} T^\mu{}_\mu T^\nu{}_\nu
}= \pm \,\tau \dot\ell(\tau).
\end{eqnarray}
The root flow equation for each theory is given by:
\begin{eqnarray}\label{ffgdgf}
  	\frac{\partial \mathcal{L}}{\partial \gamma}=\tau \dot\ell(\tau).
\end{eqnarray}
To study the root flow equation \eqref{ffgdgf} for  $q$-deformed theory, we differentiate its left-hand side with respect to the Lagrangian, yielding:
    \begin{eqnarray}\label{rfe21}
  	\frac{\partial \mathcal{L}_{q=3/4}}{\partial \gamma}&=& \frac{e^{-\gamma}\Lambda U-\frac{2}{3}e^{-2\gamma}\lambda U^2+e^{\gamma}V}{\sqrt{\Lambda-\frac{2}{3}e^{-\gamma}\lambda U}}.
\end{eqnarray}
For the right-hand side, we obtain the average root flow from equations \eqref{qdeflll} in $q=3/4$ and \eqref{tau2}:
 \begin{eqnarray}\label{rfe22}
  	\tau \dot\ell(\tau)&=& \frac{e^{-\gamma}\Lambda U-\frac{2}{3}e^{-2\gamma}\lambda U^2+e^{\gamma}V}{\sqrt{\Lambda-\frac{2}{3}e^{-\gamma}\lambda U}}.\end{eqnarray}
    A comparison of \eqref{rfe21} and \eqref{rfe22} allows us to derive the root flow equation corresponding to the $\frac{3}{4}\text{-deform}$ theory with Lagrangian \eqref{L2}:
    \begin{eqnarray}
  	\frac{\partial \mathcal{L}_{q=3/4}}{\partial \gamma}=\tau \dot\ell(\tau).
\end{eqnarray}
\subsection*{$\varphi$-parity:}
It can be shown both explicitly and perturbatively that in the $\gamma = 0$ case, the $q = 3/4$-deform  Lagrangian in~\eqref{L2} is not invariant under the transformation $(U, V) \to (-V, -U)$ and thus violates $\varphi$-parity~\cite{Russo:2025fuc}. Consequently, we expect that $\varphi$-parity is also broken in the general case where $\gamma$ is present.  
The $q = 3/4$-deform theory described in Eq.~\eqref{L2} can be obtained from the Russo--Townsend framework by considering the $\Omega(y)$ potential with an additional $\gamma$-coupling, starting from the following form~\cite{Babaei-Aghbolagh:2025uoz}:
  \begin{equation}
 \Omega(y)=\frac{1}{4\lambda} (\frac{e^{3 \gamma}}{y^3} + \frac{3 y}{e^{\gamma}}-4 )\,,
\end{equation}
Using this potential, we can systematically analyze the $\varphi$-parity properties of the $q = 3/4$-deform theory. We find that the theory is {\it not invariant} under the transformations given in Eq.~\eqref{pariii}, indicating explicit {\it$\varphi$-parity violation}. In other words, the $q = 3/4$-deform theory breaks $\varphi$-parity symmetry.

\subsection*{Irrelevant flow equation:}
Since the $q=3/4$-deform theory is not invariant under $\varphi$-parity, the irrelevant $T\bar{T}$-deformation of this theory is predicted to belong to the class of second-type irrelevant deformations, characterized by half-integer powers of the two structures  $\left(T_{\mu\nu}T^{\mu\nu}\right)$ and $\left({T_{\mu}}^{\mu} {T_{\nu}}^{\nu}\right)$, as noted in \eqref{nonAN}. In this section, we explicitly prove this claim for the $q = 3/4$ theory given in \eqref{L2}. To do so, we first compute the derivative of the Lagrangian \eqref{L2} with respect to $(\lambda)$, which yields the following expression:
\begin{eqnarray}\label{ife2}
  	\frac{\partial \mathcal{L}_{q=3/4}}{\partial \lambda}&=&\\&-& \frac{e^{-\frac{3\gamma}{2}}(-9\sqrt{3}e^{2\gamma}\Lambda^2+3\sqrt{3}e^\gamma \lambda \Lambda U+2\sqrt{3}\lambda^2 U^2+9e^{\frac{3\gamma}{2}}\sqrt{3e^\gamma \Lambda-2\lambda U}-9\sqrt{3}e^{3\gamma}\lambda V)}{9\lambda^2\sqrt{3e^\gamma \Lambda-2\lambda U}}.\nonumber
\end{eqnarray}
Building on the result that the $q = \frac{3}{4}$ theory belongs to the class of second-type irrelevant $T \bar{T}$-like deformations, we proceed by explicitly identifying the two structures in~\eqref{ttg}, which are derived from the Lagrangian in \eqref{L2}. The expressions for these two structures are as follows:
\begin{eqnarray}\label{TT2}
   T_{\mu\nu}T^{\mu\nu}&=&4 \Bigg(\frac{\Big(e^{-\gamma} U(\Lambda-\frac{2}{3}e^{-\gamma} \lambda U)+e^\gamma V\Big)^2}{\Lambda - \frac{2}{3}e^{-\gamma}\lambda U}\nonumber\\&+&\Big(\frac{1}{\lambda}-\frac{(\Lambda-\frac{2}{3}e^{-\gamma} \lambda U)^{\frac{3}{2}}}{\lambda}-\frac{e^{-\gamma} U(\Lambda-\frac{2}{3}e^{-\gamma} \lambda U)+e^\gamma V}{ \sqrt{\Lambda - \frac{2}{3}e^{-\gamma}\lambda U}}\Big)^2\Bigg),
   \end{eqnarray}
   and
  \begin{eqnarray}\label{TrTr2}
{{T_\mu}}^\mu{T_\nu}^\nu&=&16\Big(\frac{1}{\lambda}-\frac{(\Lambda-\frac{2}{3}e^{-\gamma} \lambda U)^{\frac{3}{2}}}{\lambda}-\frac{e^{-\gamma} U(\Lambda-\frac{2}{3}e^{-\gamma} \lambda U)+e^\gamma V}{ \sqrt{\Lambda - \frac{2}{3}e^{-\gamma}\lambda U}}\Big)^2.
\end{eqnarray} 
We derive the irrelevant $T\bar{T}$-like flow equation for  $q=\frac{3}{4}$ deformed theory by comparing the expansions of Eqs.~\eqref{ife2}, \eqref{TT2}, and \eqref{TrTr2} order by order. The resulting expression is:
\begin{eqnarray}\label{ife23}
	\frac{\partial \mathcal{L}_{q=3/4}}{\partial \lambda} &=&\frac{1}{24} X  + \frac{1}{18} \sqrt{X Y} - \frac{7}{216} Y - \frac{23 }{1296 }\frac{ Y^{3/2}}{ \sqrt{X}} +\frac{7}{1944 }\frac{Y^2}{ X} - \frac{869}{186624}\frac{Y^{5/2}}{ X^{3/2}} + \frac{155 }{46656 }\frac{Y^3}{ X^2}\nonumber\\&+&...,
\end{eqnarray}
where
 $X=T_{\mu\nu}T^{\mu\nu}$ and $Y={T_{\mu}}^{\mu}{T_{\nu}}^{\nu}$. We simplify the irrelevant flow equation of the $\frac{3}{4}\text{-deform}$  theory as follows:
 \begin{eqnarray}\label{irrq34}
    \frac{\partial \mathcal{L}_{q=3/4}}{\partial \lambda} &=&\sum_{m=0}^{\infty} C_m Y^{\frac{m}{2}}X^{1-\frac{m}{2}},
 \end{eqnarray}
 with the constants $C_m$
  directly obtained from irrelevant $T\bar{T}$-like flow equation \eqref{ife23}.
  This model represents the first example we have presented of a closed-form, non-analytic theory involving irrelevant $T \bar{T}$-like deformations that take the form of half-integer powers of two structures $\left(T_{\mu\nu}T^{\mu\nu}\right)$ and $\left({T_{\mu}}^{\mu} {T_{\nu}}^{\nu}\right)$.~\footnote{We find that to all orders in the deformed theory, the single trace of the energy–momentum tensor satisfies the equation
$
\frac{\partial \mathcal{L}_{q=3/4}}{\partial \lambda} = -\frac{1}{4\lambda}{T_\mu}^{\mu},
$
which can be identified as the exact renormalization-group equation for the theory. This single-trace flow equation is valid within any deformation theory, as examined in the remainder of this paper.} We will examine another instance of this type theories later in the paper.
\subsection{No $\tau$-maximum  Theory }
Since $\tau = V$ when $U = 0$, the maximum value of $\tau$ allowed by the reality of $\ell(\tau)$ also corresponds to the maximum allowed value of $V$ for the reality of the general Born-Infeld Lagrangian density.
In contrast, the theory investigated in~\cite{ Russo:2024llm, Russo:2024ptw} involves a field $\tau$ that is not subject to a maximum value and respects the principle of causality. This is referred to as the ``no $\tau$-maximum '' theory. For this theory, we consider the CH function as follows:

\begin{eqnarray}\label{l1}
	\ell(\tau)=\frac{2}{3\lambda}(1+ e^\gamma \lambda \tau)^\frac{3}{2}-\frac{2}{3\lambda}.
\end{eqnarray}
 Given the CH function~\eqref{l1} and solving the second equation, \eqref{Landtau}, we obtain $\tau$ as follows:
  \begin{eqnarray}\label{tau1}
    \tau=\frac{e^{-\gamma}(\bigtriangleup+e^\gamma \lambda V-1)}{2\lambda},
\end{eqnarray}
where $ \bigtriangleup=\sqrt{(1+\lambda V)^2+4 \lambda \,e^{-2\gamma}U}$.
Substituting the expression for $\tau$ from \eqref{tau1} into the Lagrangian \eqref{Landtau}, we obtain the Lagrangian for the ``no $\tau$-maximum'' theory in the form:
    \begin{eqnarray}\label{L1}
	{\cal L}_{No-Max}=
    \frac{2}{3 \lambda}\Big( \sqrt{2}\, (1 + 
   e^\gamma \lambda V - \frac{\bigtriangleup}{2}) \sqrt{
 \bigtriangleup +e^\gamma \lambda V+1 }\Big).
   \end{eqnarray}
The Lagrangian of no $\tau$-maximum theory in~\eqref{L1}  satisfies the duality condition given by the PDE in \eqref{sdc} and respects the principle of causality, fulfilling the criteria in  \eqref{cc} provided that $\gamma\ge0$.
It can be explicitly shown from the Lagrangian expansion in Eq.~\eqref{L1} that this theory is not invariant under $\varphi$-parity. Consequently, we expect the corresponding irrelevant deformation to involve half-integer powers of the two structures $X$ and $Y$.
\subsection*{Root flow equation:}
In the presence of $\gamma$-coupling for deformed theories, the root-$T \bar{T}$ flow equation given in Eq.~\eqref{ffgdgf} holds. For  theory with no $\tau$-maximum, we derive the left-hand side root-$T \bar{T}$ flow equation ~\eqref{ffgdgf} by taking the derivative of the Lagrangian in Eq.~\eqref{L1} with respect to $\gamma$, yielding:
 \begin{eqnarray}\label{rfe11}
  	\frac{\partial \mathcal{L}_{No-Max}}{\partial \gamma}&=& \frac{(\bigtriangleup+e^\gamma \lambda V-1)\sqrt{\bigtriangleup+e^\gamma \lambda V+1}}{2\sqrt{2}\lambda}.
\end{eqnarray}
Using Eqs. \eqref{l1} and \eqref{tau1}, the left-hand side of the root-$T \bar{T}$ flow equation \eqref{ffgdgf} can be derived as follows:
\begin{eqnarray}\label{rfe12}
  	\tau \dot\ell(\tau)&=& \frac{(\bigtriangleup+e^\gamma \lambda V-1)\sqrt{\bigtriangleup+e^\gamma \lambda V+1}}{2\sqrt{2}\lambda}.\end{eqnarray}
    By comparing \eqref{rfe11} and \eqref{rfe12}, we establish the root-$T \bar{T}$ flow equation for the no $\tau$-maximum  theory described by the Lagrangian in \eqref{L1}, as follows:
    \begin{eqnarray}
  	\frac{\partial \mathcal{L}_{No-Max}}{\partial \gamma}=\tau \dot\ell(\tau).
\end{eqnarray}
Employing the general approach outlined in \cite{ Babaei-Aghbolagh:2025uoz}, we incorporate $\gamma$-coupling into all theories studied in this paper, both in closed and perturbative form, thereby ensuring the validity of the root flow equation \eqref{ffgdgf}. For the perturbative theories discussed subsequently, we assume that the flow equation \eqref{ffgdgf} holds, though a detailed analysis is beyond the scope of this work.
\subsection*{Irrelevant flow equation:}
Since the Lagrangian in Eq.~\eqref{L1} is not invariant under $\varphi$-parity, we expect the irrelevant flow equations for the no $\tau$-maximum theory to be consistent with the irrelevant operator given in \eqref{nonAN}. To demonstrate this, we compute the derivative of the Lagrangian \eqref{L1} with respect to the coupling constant $\lambda$ as follows:
\begin{eqnarray}\label{ife1}
  	\frac{\partial \mathcal{L}_{No-Max}}{\partial \lambda}&=& \frac{8+\sqrt{2}\big(\bigtriangleup+e^\gamma \lambda V-5\big)\sqrt{\bigtriangleup+e^\gamma \lambda V+1}}{12\lambda^2}. 
\end{eqnarray}
Furthermore, the two independent energy-momentum tensor structures presented in \eqref{ttg} can be explicitly computed for the Lagrangian given in \eqref{L1}. Consequently, the no $\tau$-maximum theory described in \eqref{L1} is derived as follows:
\begin{eqnarray}\label{TT1}
   T_{\mu\nu}T^{\mu\nu}&=&\frac{4}{9}\Big(\frac{6}{\lambda^2}-\frac{9e^{-\gamma}U}{\lambda}+5e^{3\gamma}\lambda V^3+3V(5U+e^{2\gamma}V)\Big)\nonumber\\&+&\frac{4\bigtriangleup e^{-\gamma}}{9\lambda^2}(2e^\gamma+5\lambda U-2e^{2\gamma} \lambda V+5e^{3\gamma}\lambda^2 V^2)\nonumber\\&+&\frac{4\sqrt{2}\sqrt{\bigtriangleup+e^\gamma \lambda V+1}}{9\lambda^2}(\bigtriangleup+e^\gamma \lambda V-5),
   \end{eqnarray}		
and
\begin{eqnarray}\label{TrTr1}
{{T_\mu}}^\mu{T_\nu}^\nu&=&\frac{\Big(8+\sqrt{2}\,\sqrt{\bigtriangleup+e^\gamma \lambda V+1}\,(\bigtriangleup+e^\gamma \lambda V-5)\Big)^2}{9\lambda^2}.
\end{eqnarray}
 Expanding \eqref{TT1} and \eqref{TrTr1} order by order and comparing the result with the $\lambda^n$ expansion in \eqref{ife1} allows one to derive the irrelevant flow equation for no $\tau$-maximum theory.
The irrelevant flow equation for  no $\tau$-maximum theory is obtained as follows:
\begin{eqnarray}\label{ife13}
	\frac{\partial \mathcal{L}_{No-Max}}{\partial \lambda} &=&\frac{1}{16} X  -  \frac{1}{6} \sqrt{X Y} + \frac{25}{288} Y + \frac{11 }{432 }\frac{ Y^{3/2}}{ \sqrt{X}} \nonumber\\&+& \frac{1}{324 }\frac{Y^2}{ X} + \frac{269}{62208 }\frac{Y^{5/2}}{ X^{3/2}} + \frac{23 }{7776 }\frac{Y^3}{ X^2}+...\,.
\end{eqnarray}
 The irrelevant flow equation of the no $\tau$-maximum  theory can be simplified as follows:
 \begin{eqnarray}
    \frac{\partial \mathcal{L}_{No-Max}}{\partial \lambda} &=&\sum_{m=0}^{\infty} C_m Y^{\frac{m}{2}}X^{1-\frac{m}{2}},
 \end{eqnarray}
 where the constants $C_m$
are explicitly determined from equation \eqref{ife13} of the theory. 
In this section, we have demonstrated through clear examples that $\varphi$-parity-violating theories can arise from irrelevant deformations constructed from half-integer powers of the two structures, $\left(T_{\mu\nu}T^{\mu\nu}\right)$ and $\left({T_{\mu}}^{\mu} {T_{\nu}}^{\nu}\right)$. However, these examples do not constitute a general proof. Further evidence and a rigorous argument are required to establish the claim for all such theories. In the following section, we will provide a general proof using a perturbative approach, thereby confirming this conjecture.
 \section{Perturbative models: irrelevant deformations and $\varphi$-parity }\label{0044}
 A deep understanding of electromagnetic theories can be attained through the perturbation approach to dual invariant theories, which offers a unified perspective applicable to all such models. Although this approach is perturbative, the analysis is carried out with unfixed coefficients, allowing the results to be generalized to all deformed theories. The behavior of a specific theory can then be understood by fixing those coefficients accordingly.

In this section, we study $q$-deformed theories without fixing the parameter $q$ to any particular value, and we investigate their behavior under $\varphi$-parity transformations. We also identify the irrelevant deformation corresponding to this perturbative framework and examine its relationship to $\varphi$-parity. 

Furthermore, we consider a generalization of the CH function with unfixed coefficients. This extension yields a master theory that encompasses all theories discussed in the literature. We refer to this as \textbf{ \it Courant-Hilbert perturbation theory} and denote it by $\mathcal{L}_{CH}$.
We examine the invariance of this theory under $\varphi$-parity transformations as expressed in Equation~\eqref{pariii}, which leads to nontrivial conditions among the coefficients. These conditions enable the construction of a perturbative theory that remains invariant under $\varphi$-parity. By studying the irrelevant deformations of this perturbation theory, we prove that the conditions derived from $\varphi$-parity invariance eliminate half-integer powers of the two structures $\left(T_{\mu\nu}T^{\mu\nu}\right)$ and $\left({T_{\mu}}^{\mu} {T_{\nu}}^{\nu}\right)$ in the irrelevant deformation consistent with Courant-Hilbert perturbation theory. This provides a proof that analytic theories are compatible only with deformations containing integer powers of $\left(T_{\mu\nu}T^{\mu\nu}\right)$ and $\left({T_{\mu}}^{\mu} {T_{\nu}}^{\nu}\right)$.
\subsection{$q$-deformed theories}\label{4411}
Consider the (CH) function of the $q$-deformed theory given in Eq.~\eqref{qdeflll}. We have previously shown for the special case $q = 3/4$ that the function is not invariant under the $\varphi$-parity transformation. Furthermore, the corresponding irrelevant deformation in Eq.~\eqref{irrq34} takes the form of a half-integer power in $\left(T_{\mu\nu}T^{\mu\nu}\right)$ and $\left({T_{\mu}}^{\mu} {T_{\nu}}^{\nu}\right)$.
In this section, we examine the general case for this class of theories. The Lagrangian for the perturbed theory is obtained by expanding the CH function in Eq.~\eqref{qdeflll}, whose expansion we derive as follows:
\begin{eqnarray}\label{CHfq}
\ell(\tau) &=& e^\gamma \tau+\sum_{i=1}^{\infty}  \frac{(-1)^i \,\, q!}{( i+1)! \,\, q^{ i+1}\,\,  ( q-i-1)!} \, \lambda^i \, e^{( i+1) \gamma}\, \tau^{ i+1}\,.
\end{eqnarray}
In the limit $\lambda \to 0$, the CH function expansion of the $q$-deformed theory in Eq.~\eqref{CHfg} reduces to that of the ModMax theory. In this section, we consider the expansion of Eq.~\eqref{CHfg} up to order $\lambda^6$. The value of $\tau$ to order in $\lambda^6$ can be determined analytically by solving the second equation in \eqref{Landtau} for variables $U$ and $V$. From the expansion of $\tau$ to order $\lambda^6$, we derive the corresponding Lagrangian up to $\mathcal{O}(\lambda^6)$ via perturbation theory, as follows:
\begin{eqnarray}\label{qlag}
   \mathcal{L}_q&=& -e^{-\gamma} U + e^{\gamma} V \\&-& (
 \lambda e^{-2 \gamma} (q-1) (U + 
    e^{2 \gamma} V)^2)/(2 q) \nonumber\\&+& (
 \lambda^2 e^{-3 \gamma} (q-1) (U + 
    e^{2 \gamma} V)^2 ((4 - 5 q) U + 
    e^{2 \gamma} (q-2) V))/(6 q^2) \nonumber\\&-& (
 \lambda^3 e^{-4 \gamma} (q-1) (U + 
    e^{2 \gamma} V)^2 (( 7 q-6) (7 q-5) U^2 + 
    2 e^{2 \gamma} ( q (7 + q)-6) U V\nonumber\\&& 
    +e^{4 \gamma} (q-3) (q-2) V^2))/(24 q^3) \nonumber\\&+ &
  ( \lambda^4 e^{-5 \gamma} (q-1) (U + 
     e^{2 \gamma}
       V)^2 (-3 ( 3 q-2) ( 9 q-8) (9 q-7) U^3\nonumber\\& &
    - e^{2 \gamma} (72 + 
        q (-38 + q ( 257 q-273 ))) U^2 V - 
     e^{4 \gamma} (-48 + q (62 + q ( 7 q-3))) U
       V^2 \nonumber\\&& 
    + e^{6 \gamma} (q-4) (q-3) (q-2) V^3)) /(120 q^4)\nonumber\\& - &
   (\lambda^5 e^{-6 \gamma} (q-1) (U + 
     e^{2 \gamma}
       V)^2 ((11 q-10) ( 11 q-9) ( 11 q-8) (11 q-7) U^4 \nonumber\\&& 
     +4 e^{2 \gamma}
       q (-1114 + q (4421 + q (2521 q-5804 ))) U^3 V\nonumber\\& & 
    + 6 e^{4 \gamma} (60 + 
        q (-74 + q(-9 + q ( 201 q-154)))) U^2 V^2\nonumber\\&  &
     +4 e^{6 \gamma} (-60 + 
        q (86 + q (-19 + q (16 + q)))) U V^3\nonumber\\&  &
     +e^{8 \gamma} (q-5) (q-4) (q-3) (q-2) V^4))/(720 q^5) \nonumber\\&+ &
  ( \lambda^6 e^{-7 \gamma} (q-1) (-(( 13 q-12) (13 q-11) (13 q-10) (
          13 q-9) (13 q-8) U^7)\nonumber\\&  &
     -7 e^{2 \gamma} (11 q-10) ( 11 q-9) (11 q-8) (11 q-7) ( 11 q-6) U^6 V \nonumber\\&& 
    - 63 e^{4 \gamma} ( 3 q-2) (9 q-8) ( 9 q-7) ( 9 q-5) ( 9 q-4) U^5 V^2 \nonumber\\&& 
    - 35 e^{6 \gamma} ( 7 q-6) ( 7 q-5) (7 q-4) ( 7 q-3) ( 7 q-2) U^4 V^3 \nonumber\\&& 
     -175 e^{8 \gamma}
       q ( 5 q-4) ( 5 q-3) ( 5 q-2) (5 q-1) U^3 V^4 - 
     63 e^{10 \gamma}
       q (4 - 45 q^2 + 81 q^4) U^2 V^5\nonumber\\&  &
   -  7 e^{12 \gamma}
       q (1 + q) (2 + q) (3 + q) (4 + q) U V^6 \nonumber\\& &
   +  e^{14 \gamma} (q-6) (q-5) (q-4) (q-3) (q-2) V^7))/(5040 q^6)+\mathcal{O}[\lambda]^7.\nonumber
\end{eqnarray}
We have explicitly checked that the Lagrangian \eqref{qlag} of the $q$-deformed theory is not invariant under $\varphi$-parity transformations \eqref{pariii} in the general case. This conclusion was anticipated from the known result for  special case $q = 3/4$. However, a notable exception exists for $q = 1/2$, where the Lagrangian in \eqref{qlag} should be invariant under the $\varphi$-parity transformations defined in \eqref{pariii}. This suggests that the deformation flow for a general $q$ follows an operator constructed from half-integer powers of the energy-momentum tensor. In the special case $q = 1/2$, this operator reduces to one with integer powers, analogous to the structure found in generalized Born-Infeld theory.

Using the Lagrangian \eqref{qlag}, we can derive the two composite structures $X=T_{\mu\nu}T^{\mu\nu}$ and $Y={T_{\mu}}^{\mu}{T_{\nu}}^{\nu}$ for  $q$-deformed theory. Due to their lengthy explicit forms, we present these structures in the Appendix (see \eqref{TTpq} and \eqref{TrTrpq}, respectively). To derive the associated flow equations, we first compute the derivative of the Lagrangian \eqref{qlag} with respect to the deformation parameter $\lambda$ and match it order-by-order to combinations of $X$ and $Y$. This procedure yields an irrelevant flow equation driven by half-integer powers of $X$ and $Y$ for the Lagrangian \eqref{qlag}, given by:
 \begin{eqnarray}
    \frac{\partial \mathcal{L}_q}{\partial \lambda} &=&\sum_{m=0}^{\infty} C_mY^{\frac{m}{2}}X^{1-\frac{m}{2}}.
 \end{eqnarray}
where up to $\lambda^6$ the constants $C_m$ are: 
\begin{eqnarray}\label{cmq}
    C_0&=& \frac{  1-q}{8q}, \qquad  \,\,\, \,\,\,\,\,  C_1\,= \,-\frac{ 1-2q}{12 \, q} , \qquad  \,\,\,  C_2 \,=\, -\frac{ 8 q^2+q-7}{288 \, q\, ( q-1) }, \nonumber\\
   C_3 &=& \frac{  ( 1-2q) (49 + q ( 43 q-88))}{4320 (q-1)^2 q}, \qquad \,\,\,\,\,\,\,\,\,\, C_4 =   \frac{  (1 - 2 q)^2 ( q-2) ( q+1)}{12960 ( q-1)^3 q} \nonumber\\
   C_5 &=&\frac{ (1-2q) (3499 + q (-12512 + q (16926 + q ( 2335 q-10088))))}{
 4354560 ( q-1)^4 q}.\nonumber
\end{eqnarray}
All information regarding the type of deformation and its behavior under $\varphi$-parity is encoded in the coefficients $ C_m $ of the irrelevant operator in Eq.~\eqref{cmq}. In general, these coefficients correspond to a deformation built from half-integer powers of $ X $ and $ Y $, which describes a non-analytic theory. However, for  special value $ q = 1/2 $, the coefficients adjust automatically: all odd coefficients $ C_{2k+1} $ vanish, and all even coefficients $ C_{2k} $ with $ k\geq  2 $ also vanish. As a result, the surviving coefficients $\{C_0=\frac{1}{8},\, C_1=0,\,C_2=- \frac{1}{16},\, C_3=0\,,C_4=0, \,C_5= 0\}$ match the standard deformation coefficients of the general Born-Infeld theory, which describes an analytic theory. Thus, the operator reduces to a standard analytic form in this special case. We can also verify that for $ q = 3/4 $, the coefficients $ C_m $ precisely reproduce the irrelevant deformation coefficients $\{C_0=\frac{1}{24}, C_1=\frac{1}{18}\,,C_2=- \frac{7}{216} , C_3=-\frac{23 }{1296 }\,,C_4=\frac{7}{1944 }, C_5=- \frac{869}{186624}\}$ identified in Eq.~\eqref{ife23}.

Furthermore, by substituting the relations for $\left(T_{\mu\nu}T^{\mu\nu}\right)$ and $\left({T_{\mu}}^{\mu} {T_{\nu}}^{\nu}\right)$ from Eqs.\eqref{TTpq} and \eqref{TrTrpq} into the operator \(  \mathcal{R}_\gamma \) and constructing this operator explicitly, one can compare it with the derivative of the Lagrangian with respect to the coupling constant \( (\gamma) \). This comparison demonstrates that the Lagrangian in Eq.~\eqref{qlag} indeed satisfies the flow equation \( \frac{\partial \mathcal{L}_q}{\partial \gamma}= \mathcal{R}_\gamma =\tau \dot\ell(\tau). \)
 \subsection{Courant–Hilbert perturbation theory} \label{4422} 
In this section, we construct a general perturbative extension of the Courant-Hilbert function $\ell(\tau)$ up to order $\lambda^6$.  We begin by proposing a general ansatz for the CH function of the following form:
\begin{eqnarray}\label{CHftg}
	\ell(\tau)= e^\gamma \tau+\sum_{i=1}^{\infty}m_i\lambda^ie^{(i+1)\gamma}\tau^{i+1},
\end{eqnarray}
Using this expanded functional form, we solve the second equation in \eqref{Landtau} perturbatively to determine $\tau$. From this solution, we derive the corresponding general theory at $\mathcal{O}(\lambda^6)$, which we denote by $\mathcal{L}_{\text{CH}}$. The Lagrangian of the \textbf{ \it Courant-Hilbert perturbation theory} takes the following form:
\begin{eqnarray}\label{CHfg}
\mathcal{L}_{CH}& =&-e^{-\gamma}U+ e^\gamma V + e^{-2\gamma}  \lambda m_1 ( U +  e^{2\gamma} V)^2 + e^{-3 \gamma} \lambda^2 (U + e^{2 \gamma} V)^2 (-4 m_1^2 U +  m_2 (U + e^{2 \gamma} V))\nonumber\\
&+& e^{-4 \gamma} \lambda^3 (U +  e^{2 \gamma}  V)^2 (-12 m_1 m_2 U (U + e^{2 \gamma} V) + m_3 (U + e^{2 \gamma} V)^2 + 8 m_1^3 U (3 U + e^{2 \gamma} V))\nonumber\\
&+&e^{-5 \gamma}\lambda^4(U + e^{2 \gamma} V)^2 \Big(-9 m_2^2 U (U + e^{2 \gamma} V)^2 -  16 m_1 m_3 U (U + e^{2 \gamma} V)^2 + m_4 (U + e^{2 \gamma}V)^3 \nonumber\\
&+& 12 m_1^2 m_2 U (U + e^{2 \gamma} V) (11 U +3 e^{2 \gamma} V) -16 m_1^4 U (11 U^2 +8 e^{2 \gamma} U V +e^{4 \gamma} V^2)\Big)\nonumber\\
&+& e^{-6 \gamma}\lambda^5(U+ e^{2 \gamma} V)^2\Big(-24 m_2 m_3 U (U+ e^{2\gamma} V)^3 + m_5 (U + e^{2 \gamma} V)^4\\
&+&16 m_1^2 m_3 U (U + e^{2 \gamma} V)^2 (13 U+3 e^{2 \gamma} V) -  16m_1^3 m_2 U (U + e^{2 \gamma} V)\nonumber\\
&\times&(91 U^2 + 57 e^{2 \gamma} U V +  6 e^{4 \gamma} V^2)+ 16 m_1^5 U (91 U^3 +104 e^{2 \gamma} U^2 V +31 e^{4 \gamma} U V^2 + 2 e^{6 \gamma} V^3)\nonumber\\
&+&m_1 (-20 m_4 U (U+ e^{2 \gamma} V)^3 + 18 m_2^2 U (U + e^{2 \gamma} V)^2 (13 U +  3 e^{2 \gamma} V))\Big)\nonumber\\
&+&e^{-7 \gamma}\lambda^6 (U + e^{2 \gamma} V)^2 \Big(-16 m_3^2 U (U + e^{2 \gamma} V)^4 -  30 m_2 m_4 U (U + e^{2 \gamma} V)^4 \nonumber\\
&+&m_6 (U + e^{2 \gamma} V)^5 +  27m_2^3 U (U + e^{2 \gamma} V)^3 (5 U +  e^{2 \gamma} V)\nonumber\\
&-&128 m_1^3 m_3 U (U + e^{2 \gamma} V)^2 (20 U^2 + 11 e^{2 \gamma} U V + e^{4 \gamma} V^2) \nonumber\\
&+&240m_1^4 m_2 U(U + e^{2 \gamma} V)(68 U^3 + 69 e^{2 \gamma} U^2 V + 18 e^{4 \gamma} U V^2 + e^{6 \gamma} V^3) \nonumber\\
&-& 64m_1^6 U (204 U^4 + 320 e^{2 \gamma} U^3 V + 157 e^{4 \gamma} U^2 V^2 + 26 e^{6 \gamma} U V^3 + e^{8 \gamma} V^4) \nonumber\\
&+& m_1 (-24 m_5 U (U + e^{2 \gamma} V)^4 +   144 m_2 m_3 U (U + e^{2\gamma}V)^3 (5 U + e^{2 \gamma} V)) \nonumber\\
&+& m_1^2 (60  m_4 U (U + e^{2 \gamma} V)^3 (5 U +  e^{2 \gamma}V)-216m_2^2 U (U + e^{2 \gamma} V)^2 (20 U^2+ 11 e^{2 \gamma}U V +e^{4 \gamma} V^2))\Big).\nonumber
\end{eqnarray}
The Courant–Hilbert perturbation theory is a general theory that is not invariant under the $\varphi$-parity transformations in \eqref{pariii}. Consequently, we expect its irrelevant deformation to be governed by an operator built from half-integer powers of the two independent energy-momentum tensor structures.
For  Lagrangian in \eqref{CHfg}, we compute the energy-momentum tensor and extract the two independent scalar structures, \( T_{\mu\nu}T^{\mu\nu} \) and \( ({T^\mu}_\mu)^2 \), up to order \(\lambda^6\). Their explicit forms are given in Eqs.~\eqref{TTg2} and \eqref{TrTrg}.
By comparing the derivative of the Lagrangian~\eqref{CHfg} with respect to \(\lambda\) with combinations of these two structures, we obtain the irrelevant flow equation governing the general perturbative theory:
 \begin{eqnarray}\label{CHIrr}
    \frac{\partial \mathcal{L}_{CH}}{\partial \lambda} &=&\sum_{m=0}^{\infty} C_m Y^{\frac{m}{2}}X^{1-\frac{m}{2}}.
 \end{eqnarray}
where  $X=T_{\mu\nu}T^{\mu\nu}$, $Y={T_{\mu}}^{\mu}{T_{\nu}}^{\nu}$ and the constants $C_m$
are derived as follows:
\begin{eqnarray}\label{cm}
   {C_0}&=&\frac{m_1}{4}\nonumber\\C_1&=&\frac{-2 m_1^2 + m_2}{4 m_1}\nonumber\\C_2&=&\frac{3 m_1^4 - 2 m_1^2 m_2 - 4 m_2^2 + 3 m_1 m_3}{16 m_1^3}\nonumber\\C_3&=&\frac{2 m_1^6 - m_1^4 m_2 + 16 m_2^3 + 4 m_1^3 m_3 - 18 m_1 m_2 m_3 + 
 m_1^2 (-6 m_2^2 + 4 m_4)}{ 32 m_1^5}\nonumber\\
 C_4 & =& \frac{-80 m_2^4 + 8 m_1^5 m_3 + 120 m_1 m_2^2 m_3}{64 m_1^7}\nonumber\\& +& 
   \frac{m_1^4 (-13 m_2^2 + 14 m_4) + 2 m_1^2 (32 m_2^3 - 9 m_3^2 - 16 m_2 m_4) + 
   m_1^3 (-68 m_2 m_3 + 5 m_5)}{64 m_1^7} \nonumber\\
 C_5 & =& \frac{1}{ 512 m_1^9} \Big(2 m_1^{10} - m_1^8 m_2 + 1792 m_2^5 + 72 m_1^7 m_3 - 
     3360 m_1 m_2^3 m_3 \nonumber\\&+& 4 m_1^6 (-31 m_2^2 + 42 m_4) +
     40 m_1^2 m_2 (-52 m_2^3 + 27 m_3^2 + 24 m_2 m_4)\nonumber\\& + &
     28 m_1^5 (-31 m_2 m_3 + 4 m_5) + 
     40 m_1^3 (75 m_2^2 m_3 - 6 m_3 m_4 - 5 m_2 m_5) \nonumber\\&+& 
     8 m_1^4 (107 m_2^3 - 54 m_3^2 - 95 m_2 m_4 + 3 m_6)\Big).
\end{eqnarray}
Based on our earlier observations and the conjecture introduced in this work, that analytic theories of irrelevant deformations arise from integer powers of the two independent energy-momentum tensor structures, the requirement for analytic theories up to order $\lambda^6$ imposes that all odd coefficients as $C_1$, $C_3$, and $C_5,...$ must vanish. Enforcing this condition leads to the following set of constraints among the coefficients $m_i$ of the CH  function:
 \begin{eqnarray}\label{mmm333}
&&m_2= 2 m_1^2, \qquad m_4= -2 (13 m_1^4-4 m_1 m_3),\\
&&m_6=4 (357 m_1^6-100 m_1^3 m_3+2 m_3^2+3 m_1 m_5), \nonumber
 \end{eqnarray}
Therefore, if condition~\eqref{mmm333} is satisfied among the expansion coefficients of the CH function $\ell(\tau)$, then the resulting theory at order $\mathcal{O}(\lambda^6)$ will be analytic. This analytic theory is described by the Lagrangian:
  \begin{eqnarray}\label{Lparityinv}
   \mathcal{L}_{\varphi-inv}&=&- e^{-\gamma} U + e^{\gamma} V+ \lambda \, e^{-2 \gamma} m_1 (U + e^{2 \gamma} V)^2+2 \lambda^2 e^{-3 \gamma} m_1^2 (- U + e^{2 \gamma} V) (U + e^{2 \gamma} V)^2 \nonumber\\
   &+& \lambda^3 \Bigl(-16 e^{-2 \gamma} m_1^3 U V (U + e^{2 \gamma} V)^2 + e^{-4 \gamma} m_3 (U + e^{2 \gamma} V)^4\Bigr) \nonumber\\
   &+& \lambda^4 \Bigl( 2 e^{-5 \gamma} m_1^4 (U + e^{2 \gamma} V)^2 (13 U^3 + 29 e^{2 \gamma} U^2 V - 29 e^{4 \gamma} U V^2 - 13 e^{6 \gamma} V^3)\nonumber\\
   &+&8 e^{-5 \gamma} m_1 m_3 (- U + e^{2 \gamma} V) (U + e^{2 \gamma} V)^4 \Bigr)+\lambda^5 \Bigl(-160 e^{-4 \gamma} m_1^2 m_3 U V (U + e^{2 \gamma} V)^4 \nonumber\\
   &+& e^{-6 \gamma} m_5 (U + e^{2 \gamma} V)^6 + 64 e^{-4 \gamma} m_1^5 U V (U + e^{2 \gamma} V)^2 (9 U^2 + 22 e^{2 \gamma} U V + 9 e^{4 \gamma} V^2)\Bigr) \nonumber\\
   &+& \lambda^6 \Bigl(8 e^{-7 \gamma} m_3^2 (- U + e^{2 \gamma} V) (U + e^{2 \gamma} V)^6 + 12 e^{-7 \gamma} m_1 m_5 (- U + e^{2 \gamma} V) (U + e^{2 \gamma} V)^6 \nonumber\\
   &+& 80 e^{-7 \gamma} m_1^3 m_3 (U + e^{2 \gamma} V)^4 (5 U^3 + 13 e^{2 \gamma} U^2 V - 13 e^{4 \gamma} U V^2 - 5 e^{6 \gamma} V^3)\nonumber \\
   &+& 4 e^{-7 \gamma} m_1^6 (U + e^{2 \gamma} V)^2 (-357 U^5 - 1727 e^{2 \gamma} U^4 V - 1498 e^{4 \gamma} U^3 V^2 \nonumber\\
   &+& 1498 e^{6 \gamma} U^2 V^3 + 1727 e^{8 \gamma} U V^4 + 357 e^{10 \gamma} V^5)\Bigr)
 \end{eqnarray}
 Applying condition~\eqref{mmm333} to the $C_m$ coefficients defined in~\eqref{cm} forces all odd coefficients to vanish, leaving only the even coefficients. The surviving even coefficients take the following form:
\begin{eqnarray}\label{cmanaaa}
   && C_0=\frac{m_1}{4}, \quad   C_2 \,=\,- \frac{17}{16} m_1 + \frac{3 m_3}{16 m_1^2}, \quad  C_4 =  \frac{480 m_1^6 - 48 m_1^3 m_3 - 18 m_3^2 + 5 m_1 m5}{64 m_1^5} \nonumber\\
  &&C_1\,= \,0, \quad  C_3 = 0, \quad C_5 =0
\end{eqnarray}
Therefore, the analytic Lagrangian \eqref{Lparityinv} corresponds to the irrelevant deformation \eqref{CHIrr}. In this case, condition \eqref{cmanaaa} enforces that all odd coefficients vanish, so the associated deformation operator contains only integer powers of the two structures \(X\) and \(Y\). 
{ Compatibility with exact $\varphi$-parity in~\eqref{exaparit} or analytic Lagrangian as \eqref{uvuv} 
requires that \(\gamma = 0\), in which case the root-\(T\bar{T}\) flow equation is lost for deformed theories. In other words, 
the existence of an analytic theory entails the loss of the root flow 
equation in the deformed theory.}
In Ref.~\cite{Babaei-Aghbolagh:2024uqp}, a family of perturbative theories is presented whose deformations involve integer powers of the two structures of energy–momentum tensor. Consequently, these theories belong to the class of analytical theories and are invariant under $\varphi$-parity  transformation in~\eqref{pariii}. A comparison between the analytic Lagrangian \eqref{Lparityinv} and the general Lagrangian \eqref{CHfg} reveals a key structural distinction. In the analytic case, each pair of consecutive orders in $(\lambda)$ is governed by a single free coefficient $m_i$, whereas in non-analytic theories, every order in $(\lambda)$ requires an independent coefficient $m_i$ to be fixed. Finally, it should be noted that Courant–Hilbert perturbation theory applies to the root flow equation in the form:  \( \frac{\partial \mathcal{L}_{CH}}{\partial \gamma}= \mathcal{R}_\gamma =\tau \dot\ell(\tau). \)

{
Starting from a general potential \( W(\varphi) \) expanded in powers of \( \varphi \):
\begin{equation} \label{cmera}
W(\varphi) = \frac{1}{\lambda} \sum_{i=0}^{\infty} a_i \, \varphi^{i},
\end{equation}
where the coefficients \( a_i \) are constants, a perturbation Lagrangian can be generated using the new auxiliary field approach. When all odd coefficients vanish (\( a_{\text{2k+1}} = 0 \)), an analytic theory emerges in the weak-field regime up to order \( (\lambda^i )\), analogous to \eqref{Lparityinv}, involving only the even coefficients \(( a_{\text{2k}} )\) from  terms $\varphi^{2k}$.
Applying the transformation \eqref{eq:legendre} then yields a relation between the coefficients \( ( m_i )\) in \eqref{CHftg} and the coefficients \(( a_i) \) in \eqref{cmera} in the limit \( \gamma = 0 \).
}

As a result, we have introduced a framework that determines whether a theory respects the $\varphi$-parity transformation in \eqref{pariii} directly from the expansion coefficients of the CH function $\ell(\tau)$, without requiring an explicit Lagrangian formulation. Our findings show that a theory is $\varphi$-parity-invariant if the coefficients of its CH function expansion satisfy condition \eqref{mmm333}. This condition follows from the requirement that the irrelevant deformation operator for an analytic theory must be composed entirely of integer powers of the two independent energy-momentum tensor structures.

  \section{Conclusion and outlook}\label{0055}

In this work, we establish a fundamental connection among the analytic structure of self-dual nonlinear electrodynamic (NED) theories, their behavior under a discrete $\varphi$-parity transformation, and the specific form of the irrelevant $T\bar{T}$-like deformations that generate them.

Self-dual NED theories are, by construction, functions of the Lorentz invariants $S$ and $P^2$. However, they typically involve non-analytic square-root structures, such as $\sqrt{S^2+P^2}$, in both their Lagrangians and their weak-field expansions. A central question addressed in this work is identifying the general class of self-dual theories that possess a genuinely analytic weak-field expansion that is, expansions purely in powers of $S$ and $P^2$, entirely free of square roots or other non-analytic functions of the fundamental invariants $F_{\mu\nu}F^{\mu\nu}$ and $F_{\mu\nu}\widetilde{F}^{\mu\nu}$.
Our central result provides a definitive answer to this question through a precise classification. \textbf{Analytic theories}, which possess a standard weak-field Taylor expansion, are exactly those invariant under the $\varphi$-parity transformation (generalized to include a marginal $\gamma$-coupling as $\mathcal{L}(U, V,\gamma) = \hat{\mathcal{L}}(-V,-U,-\gamma)$). This analyticity manifests structurally in their associated deformation flows: such theories are generated by irrelevant $T\bar{T}$-like operators built exclusively from \textbf{integer powers} of the two independent Lorentz scalars constructed from the energy-momentum tensor, $X = T_{\mu\nu}T^{\mu\nu}$ and $Y = ({T_{\mu}}^{\mu})^2$. Conversely, \textbf{non-analytic theories}, which violate $\varphi$-parity and contain square-root structures in their weak-field expansions, arise from deformation operators that necessarily involve \textbf{both integer and half-integer powers} of $X$ and $Y$. We provided a general proof of this correspondence within a comprehensive Courant-Hilbert perturbation theory framework, showing that $\varphi$-parity invariance imposes specific constraints on the expansion coefficients of the CH function $\ell(\tau)$ which systematically eliminate all half-integer powers, and thus all square-root structures, from the deformation and, consequently, from the weak-field expansion.

We validated this classification through detailed analysis of both closed-form and perturbative examples. The analytic generalized Born-Infeld theory and the non-analytic $q=3/4$-deformed and ``no $\tau$-maximum'' theories served as paradigmatic cases, for which we explicitly derived the corresponding marginal (root-$T\bar{T}$) and irrelevant flow equations. The perturbative treatment of the $q$-deformed family further illustrated how the special case $q=1/2$ restores $\varphi$-parity and collapses the deformation operator to the analytic, integer-power form, thereby eliminating square roots from the expansion.
This work clarifies the landscape of self-dual NEDs by identifying $\varphi$-parity as the underlying symmetry principle governing analyticity, which in turn dictates the permissible structures in the irrelevant deformation. The Russo-Townsend auxiliary field formalism, with its master potential $\Omega(y)$, provides a particularly transparent framework for understanding this connection, as $\varphi$-parity invariance translates directly into the simple inversion symmetry $\Omega(y, \gamma) = \hat{\Omega}(y^{-1}, -\gamma)$.
A natural question arises\footnote{We would like to thank Dmitri Sorokin for raising this question.}: do there exist $\gamma$-dependent nonlinear electrodynamic (NED) theories for which the transformation in the $\Omega$--plane,
\begin{eqnarray}
\Omega(\gamma, y) \longrightarrow \hat{\Omega}(\gamma, 1/y),
\end{eqnarray}
is both invariant and analytic, without requiring the additional transformation $\gamma \to -\gamma$?
Equivalently, can one find Lagrangians satisfying the symmetry condition
\begin{eqnarray}\label{ffflll}
L(U,V,\gamma) \;=\; L(-V,-U,\gamma) \; ?
\end{eqnarray}
{As argued at the end of the previous section, strict parity consistency or analyticity in the Lagrangian \eqref{Lparityinv} requires \( \gamma = 0 \). In this limit, the root-$T\bar{T}$ flow equation vanishes. Without this flow, the deformed theory cannot reproduce the special duality-invariant structure characteristic of ModMax. Instead, the resulting analytic theory, although well-defined in the weak-field regime, corresponds to a different deformation path, one that does not yield ModMax, even at leading order in the fields.}
To investigate this, we consider a perturbative expansion up to order $\lambda^4$, described by a CH function of the form  
\begin{eqnarray}\label{fft}
\ell(\tau) = f_0(\gamma) \tau +\lambda m_1 f_1(\gamma) \tau^2+\lambda^2 m_2\,f_2(\gamma) \tau^3 + \lambda^3 m_3 \, f_3(\gamma) \tau^4+\lambda^4 m_4 f_4(\gamma) \tau^5 ,
\end{eqnarray}
where the $f_i(\gamma)$ are arbitrary functions of $\gamma$ and $m_i$ are constant.
 {The coupling \( \gamma \) in \eqref{fft} does not necessarily correspond to a root-$T\bar{T}$ flow equation. Consequently,  if such a Lagrangian~\eqref{ffflll} exists, it will not reduce to the ModMax theory in the weak-field limit.}
Applying condition \eqref{ffflll} to the Lagrangian derived from CH function\eqref{fft} yields the following constraints on the functions $f_i(\gamma)$:
\begin{eqnarray}
f_0(\gamma) = 1, \quad f_2(\gamma) =\frac{2 m_1^2 f^2_1(\gamma)}{m_2}\,, \quad f_4(\gamma) =- \frac{2 \Bigl(13 m_1^4 f^4_1(\gamma) - 4 m_1 m_3 f_1(\gamma) f_3(\gamma)\Bigr)}{m_4} .
\end{eqnarray}
Enforcing these conditions yields a theory that is  analytic and  invariant under~\eqref{ffflll} without being reducible under those transformations, and which explicitly retains $\gamma$-dependence. This  analytic Lagrangian is obtained as follows:
\begin{eqnarray}\label{Lana}
\mathcal{L}_{ analytic}&=&- U + V + m_1 \,  \lambda f_1(\gamma) \, (U + V)^2- 2 m_1^2\,\lambda^2 f^2_1(\gamma)\, (U -  V) (U + V)^2 \\
&+& \lambda^3 \Bigl(-16 m_1^3 U V (U + V)^2 f^3_1(\gamma) + m_3 (U + V)^4 f_3(\gamma)\Bigr) \nonumber\\
&+& \lambda^4 \Bigl(2 m_1^4 (U -  V) (U + V)^2 (13 U^2 + 42 U V + 13 V^2) f^4_1(\gamma) \nonumber\\
&-& 8 m_1 m_3 (U -  V) (U + V)^4 f_1(\gamma) f_3(\gamma)\Bigr)\,.\nonumber
\end{eqnarray}
By substituting the variables \(U\) and \(V\) from \eqref{UV} into \eqref{Lana}, we can express it in terms of the integer powers \(S\) and \(P\) as follows:
\begin{eqnarray}\label{LanaSP}
\mathcal{L}_{ analytic}&=&S + \lambda  \, m_1 f_1(\gamma) (P^2 + S^2) + 2 \, \lambda^2 \, m_1^2  f^2_1(\gamma) S (P^2 + S^2)\\
&-&  \lambda^3 (P^2 + S^2) \Bigl(4 m_1^3 f^3_1(\gamma) P^2 -  m_3 f_3(\gamma) (P^2 + S^2)\Bigr)\nonumber \\
&-&2 \,\lambda^4\, m_1  f_1(\gamma) S (P^2 + S^2) \Bigl(-4 m_3 f_3(\gamma) (P^2 + S^2) + m_1^3 f^3_1(\gamma) (17 P^2 + 13 S^2)\Bigr)\,.\nonumber
\end{eqnarray}
Notably, the root structure \(\sqrt{S^2 + P^2}\) does not appear.
This Lagrangian raises questions concerning the choice of functions \(f_i(\gamma)\) and the nature of the $T\bar{T}$ deformations associated with  the Lagrangian~\eqref{Lana} with respect to the couplings \((\gamma)\) and \((\lambda)\).

Our results open several promising directions for future investigation. The Courant-Hilbert framework developed here also provides a systematic method for constructing analytic, $\varphi$-parity-invariant Lagrangians, motivating the search for new closed-form theories beyond Born--Infeld and a more complete classification of non-analytic models organized by their fractional deformation structure. 
The Russo-Townsend auxiliary field formalism has recently been extended to supersymmetric theories in Ref.~\cite{Kuzenko:2025gvn}.
Extending the analysis to include dynamical matter fields and supersymmetric embeddings represents another important direction, as matter couplings may alter analyticity properties and supersymmetry could impose additional constraints linking $\varphi$-parity to internal symmetries. On the quantum side, understanding how these theories behave under renormalization group flow and assessing the robustness of the analytic versus non-analytic classification after quantization remain essential open problems.

The close relationship between self-duality, integrability, and deformation dynamics suggests that further enhanced integrable structures may exist~\cite{Babaei-Aghbolagh:2025hlm,Babaei-Aghbolagh:2025uoz,Babaei-Aghbolagh:2022leo,Ferko:2022cix,Borsato:2022tmu,Ferko:2025bhv,Baglioni:2025tsc,Sakamoto:2025hwi,Fukushima:2025tlj}. A deeper group-theoretic interpretation of $\varphi$-parity within larger electromagnetic duality groups invites particular exploration. Together, these directions highlight the role of discrete symmetry as a unifying principle for organizing self-dual electrodynamics and for guiding the construction of consistent deformed quantum field theories.


\section*{Acknowledgments}
We are very grateful to Bin Chen, Jue Hou, Dmitri Sorokin, Roberto Tateo, Jorge G. Russo, and Shahin Sheikh-Jabbari for their interest in this work and for the fruitful discussions.
H.B.-A. would like to express my sincere gratitude to Karapet Mkrtchyan, Neil Lambert, and Alessandro Tomasiello for the highly productive discussions during the "Journey through Modern Explorations in QFT and beyond - 2 $\otimes$25" conference in Yerevan (August 3-13).
The work of H.B.-A. was conducted as part of the PostDoc Program on {\it Exploring TT-bar Deformations: Quantum Field Theory and Applications}, sponsored by Ningbo University.  SH would like to appreciate the financial support from Ningbo University, the Max Planck Partner Group, and the National Natural Science Foundation of China (Grants Nos. 12475053, 12235016, and 12588101). This work is based upon research funded by the Iran National Science Foundation (INSF) under project No. 4047785.


\begin{appendix}
\section{Explicit forms of energy-momentum tensor structures }
In this appendix, we collect the detailed perturbative expansions of the two independent Lorentz scalars constructed from the energy-momentum tensor: $X \equiv T_{\mu\nu}T^{\mu\nu}$ and $Y \equiv  {T_\mu}^\mu{T_\nu}^\nu$. These expansions correspond to the theories analyzed in Sections~\eqref{4411} and~\eqref{4422}, namely:
\begin{itemize}
\item The general $q$-deformed theory (Eq.~\ref{qlag})
\item The master Courant-Hilbert perturbation theory (Eq.~\ref{CHfg})
\end{itemize}
\subsection*{Energy-momentum tensor structures of $q$-deformed theory:}
The structure expansion $ T_{\mu\nu}T^{\mu\nu}$  and ${T_\mu}^\mu{T_\nu}^\nu $ up to $\lambda^6$ order for  $q$-deformed theory is obtained as follows:
\begin{eqnarray}\label{TTpq}
  T_{\mu\nu}T^{\mu\nu}&=& 4 e^{-2 \gamma} ( U + e^{2 \gamma} V)^2 \nonumber\\ &+& 8 \lambda e^{-3 \gamma} (q-1)   (U + e^{2 \gamma} V)^2 ( U - e^{2 \gamma} V)/q \nonumber\\
  &+ &\lambda^2 e^{-4 \gamma} (q-1) (U + e^{2 \gamma} V)^2\nonumber\\&&\times ((-21 + 25 q)
      U^2 -14 e^{2 \gamma} (q-1) U
     V + e^{4 \gamma} (-13 + 9 q) V^2)/q^2\nonumber\\&+&
 4\lambda^3e^{-5 \gamma}(q-1)(
    U + e^{2 \gamma} V)^2 \nonumber\\&&\times((46 + q (-113 + 69 q))
      U^3 -e^{2 \gamma} (q-2) (-9 + 11 q)
     U^2 V + e^{4 \gamma} (q-2) (-9 + 11 q) U
      V^2 \nonumber\\&& -e^{6 \gamma} (q-2) (-7 + 5 q) V^3)/3q^3\nonumber\\&+& 
  \lambda^4 e^{-6 \gamma} (q-1) (
    U + e^{2 \gamma} V)^2 \nonumber\\&&\times((-3422 + q (
       12871 + q (-16054 + 6641 q)))
      U^4 \nonumber\\&& +4 e^{2 \gamma} (
     112 + q (-35 + q (-436 + 377 q)))
      U^3 V \nonumber\\&& +6 e^{4 \gamma} (q-1) (
     134 + q (-225 + 73 q)) U^2 V^2 \nonumber\\&& -4 e^{
     6 \gamma} (-160 + q (347 + q (-200 + 31 q))) U
     V^3 \nonumber\\&& +e^{8 \gamma} (q-2) (
     223 + q (-252 + 65 q)) V^4)/18 q^4\nonumber\\&+& \lambda^5 e^{-7 \gamma} (q-1) (
    U + e^{2 \gamma} V)^2 \nonumber\\&&\times((
     9192 + q (-46774 + q (88841 +q (-74654 + 23419 q))))
      U^5\nonumber\\& & +e^{2 \gamma} (
     1140 + q (-11452 + q (32503 + q (-36332 + 14213 q))))
      U^4 V\nonumber\\&& +2 e^{4 \gamma} (
     660 + q (-2634 + q (4261 + q (-3474 + 1211 q))))
      U^3 V^2\nonumber\\&& +2 e^{
      6 \gamma} (-480 + q (
       1384 + q (-1121 + q (104 + 89 q))))
      U^2 V^3 \nonumber\\&&+e^{8 \gamma} (
     720 + q (-1798 + q (1357 + q (-398 + 47 q)))) U
      V^4 \nonumber\\&& -e^{10 \gamma} (q-3) (q-2) (
    78 + q (-89 + 23 q)) V^5)/15q^5\nonumber\\&+&\lambda^6  e^{-8 \gamma} (q-1) (
    U + e^{2 \gamma} V)^2 \nonumber\\&& \times((-1462836 + q (
       9406328 +q (-24092305 + q (
           30725815 + 13 q (-1500983 + 379749 q)))))
      U^6 \nonumber\\&& +2 e^{
      2 \gamma} (-275436 + q (
       2339072 + q (-7452635 + q (
           11372685 + q (-8408329 + 2426563 q)))))
      U^5 V \nonumber\\&& +e^{
      4 \gamma} (-168780 + q (
       1104056 + q (-3243055 + q (
           5130025 + q (-4178165 + 1360719 q)))))
      U^4 V^2 \nonumber\\&& +4 e^{
      6 \gamma} (q-1) (-19500 + q (
       43204 + q (8289 + q (-72666 + 41509 q))))
      U^3 V^3 \nonumber\\&& +e^{
      8 \gamma} (-60300 + q (
       184936 + q (-168415 + q (47385 + q (-19205 + 10799 q)))))
      U^2 V^4\nonumber\\&& -2 e^{
     10 \gamma} (-21972 + q (
      60704 + q (-55045 + q (22675 + q (-4823 + 381 q))))) U
     V^5 \nonumber\\&& +e^{
      12 \gamma} (q-3) (q-2) (-4542 + q (
       6515 + 11 q (-258 + 35 q))) V^6)/720q^6+\mathcal{O}[\lambda]^7,
\end{eqnarray}
and
\begin{eqnarray}\label{TrTrpq}
  {T_\mu}^\mu{T_\nu}^\nu &=&  4\lambda^2 e^{-4 \gamma} (q-1)^2 (U + 
    e^{2 \gamma} V)^4/q^2 \nonumber\\& +& 
 16\lambda^3 e^{-5 \gamma} (q-1)^2 (U + 
    e^{2 \gamma} V)^4 ((-4 + 5 q) U - 
    e^{2 \gamma} (-2 + q) V)/(3 q^3)\nonumber\\& +& 
 2 \lambda^4 e^{-6 \gamma} (q-1)^2 (U + 
    e^{2 \gamma}
      V)^4 ((398 + q (-1013 + 641 q)) U^2\nonumber\\&&-
    2 e^{2 \gamma} (118 + q (-175 + 31 q)) U V + 
    e^{4 \gamma}(-2 + q) (-43 + 17 q) V^2)/(
 9 q^4) \nonumber\\&+& 
 4 \lambda^5e^{-7 \gamma} (q-1)^2 (U + 
    e^{2 \gamma}
      V)^4 ((-1272 + q (4918 + q (-6307 + 2683 q))) U^3\nonumber\\&& 
   + e^{2 \gamma} (684 + 
       q (-1576 + q (639 + 319 q))) U^2 V + 
    e^{4 \gamma} (-336 + q (574 + q (-201 + 29 q))) U
      V^2\nonumber\\&& 
    -e^{6 \gamma} (-3 + q) (-2 + q) (-18 + 
       7 q) V^3)/(15 q^5)\nonumber\\& +& \lambda^6
 e^{-8 \gamma} (q-1)^2 (U + 
    e^{2 \gamma}
      V)^4 \nonumber\\&&((227316 + 
       q (-1182884 + 
          q (2298061 + q (-1975594 + 634145 q)))) U^4\nonumber\\&& + 
    12 e^{2 \gamma} (-7820 + 
       q (21192 + 
          q (-2913 + q (-30618 + 20507 q)))) U^3 V \nonumber\\&&+ 
    6 e^{4 \gamma} (8564 + 
       q (-19964 + q (9581 + q (446 + 2417 q)))) U^2 V^2 \nonumber\\&&- 
    4 e^{6 \gamma} (5892 + 
       q (-11296 + q (5699 + q (-1466 + 127 q)))) U V^3\nonumber\\&& + 
    3 e^{8 \gamma} (-3 + q) (-2 + q) (394+ 
       q (-263 + 43 q)) V^4)/(1/180q^6) + \mathcal{O}[\lambda]^7.
\end{eqnarray}
\subsection*{Energy-momentum tensor structures of Courant–Hilbert perturbation theory:}
The structure expansion $ T_{\mu\nu}T^{\mu\nu}$  and ${T_\mu}^\mu{T_\nu}^\nu $ up to $\lambda^6$ order for the Courant–Hilbert perturbation theory is obtained as follows:
\begin{eqnarray}\label{TTg2}
   T_{\mu\nu}T^{\mu\nu}&=& 
       4 e^{-2 \gamma} (U + e^{2 \gamma} V)^2 \nonumber\\&+& 
 16 e^{-3 \gamma}
    \lambda m_1 (U +
     e^{2 \gamma} V)^2 (-U + e^{2 \gamma} V)\nonumber\\&+& 
 4 e^{-4 \gamma} \lambda^2 (U + 
    e^{2 \gamma}
      V)^2 ((29 m_1^2 - 6 m_2) U^2 - 14 e^{2 \gamma} m_1^2 U V + 
  e^{4 \gamma} (5 m_1^2 + 6 m_2) V^2) \nonumber\\&+& 
 32 e^{-5 \gamma}\lambda^3 (U + 
   e^{2 \gamma}
     V)^2 (-((32 m_1^3 - 14 m_1 m_2 + m_3) U^3) + 
   e^{2 \gamma} (6 m_1 m_2 - m_3) U^2 V \nonumber\\& &+
   e^{4 \gamma} (-6 m_1 m_2 + m_3) U V^2 + 
   e^{6 \gamma} (2 m_1 m_2 + m_3) V^3) \nonumber\\&+&
 4 e^{-6 \gamma}\lambda^4 (U + 
   e^{2 \gamma}
     V)^2 ((2496 m_1^4 - 1688 m_1^2 m_2 + 103 m_2^2 + 182 m_1 m_3 - 
      10 m_4) U^4 \nonumber\\&& +
   2 e^{2 \gamma} (584 m_1^4 - 732 m_1^2 m_2 + 71 m_2^2 + 
      124 m_1 m_3 - 10 m_4) U^3 V\nonumber\\& &- 
   4 e^{4 \gamma} (8 m_1^4 - 54 m_1^2 m_2 + 3 m_2^2 + 
      7 m_1 m_3) U^2 V^2\nonumber\\& & +
   2 e^{6 \gamma} (8 m_1^4 - 4 m_1^2 m_2 - 19 m_2^2 - 
      36 m_1 m_3 + 10 m_4) U V^3\nonumber\\&&+ 
   e^{8 \gamma} (13 m_2^2 + 22 m_1 m_3+ 10 m_4) V^4) \nonumber\\&-& 
16 e^{-7 \gamma}\lambda^5 (U + 
     e^{2 \gamma}
       V)^2 (4 m_1^2 m_3 U (U + 
        e^{2 \gamma} V)^2 (193 U^2 - 
        46 e^{2 \gamma} U V + 
        e^{4 \gamma} V^2)\nonumber\\&  &-
     12 m_1^3 m_2 U (U+ 
        e^{2 \gamma} V) (495 U^3 + 
        153 e^{2 \gamma} U^2 V - 
        19 e^{4 \gamma} U V^2 + 
        3 e^{6 \gamma} V^3)\nonumber\\&  &+
     m_1 (U + 
        e^{2 \gamma} V)^2 ((873 m_2^2 - 67 m_4) U^3 - 
        3 e^{2 \gamma} (66 m_2^2 + 7 m_4) U^2 V \nonumber\\& &+
        3 e^{4 \gamma} (3 m_2^2 + 13 m_4) U V^2 - 
        7 e^{6 \gamma} m_4 V^3) \nonumber\\& &+
     16 m_1^5 U (404 U^4 + 
        375 e^{2 \gamma} U^3 V + 
        67 e^{4 \gamma} U^2 V^2 + 
        e^{6 \gamma} U V^3 + 
        e^{8 \gamma} V^4)\nonumber\\& &- 
     3 (U + e^{2 \gamma} V)^3 (-m_5 U^2 + 
        e^{4 \gamma} m_5 V^2 + 
        3 m_2 m_3 (-3 U + e^{2 \gamma} V)^2))
  \nonumber\\&+& 
 4 e^{-8 \gamma} \lambda^6(U + 
    e^{2 \gamma}
      V)^2 (16 m_1^3 m_3 U (U+ 
       e^{2 \gamma}V)^2 (2941 U^3 + 
       521 e^{2 \gamma} U^2 V\nonumber\\&  &-
       165 e^{4 \gamma} U V^2 + 
       15 e^{6 \gamma} V^3) + 
    12 m_1^2 U (U + 
       e^{2 \gamma} V)^2 ((6647 m_2^2 - 422 m_4) U^3\nonumber\\&  &+
       e^{2 \gamma} (1237 m_2^2 - 286 m_4) U^2 V - 
       e^{4 \gamma} (331 m_2^2 - 134 m_4) U V^2 \nonumber\\&& +
       e^{6 \gamma} (39 m_2^2 - 2 m_4) V^3) - 
    2 m_1 (U + 
       e^{2 \gamma} V)^3 ((6116 m_2 m_3 - 185 m_5) U^3 \nonumber\\& &-
       e^{2 \gamma} (1880 m_2 m_3 + 51 m_5) U^2 V+ 
       e^{4 \gamma} (68 m_2 m_3 + 117 m_5) U V^2 - 
       17 e^{6 \gamma} m_5 V^3) \nonumber\\&&- (U + 
       e^{2 \gamma}
         V)^3 ((2304 m_2^3 - 249 m_3^2 - 466 m_2 m_4 + 14 m_6) U^3\nonumber\\& &- 
       e^{2 \gamma} (684 m_2^3 + 75 m_3^2 + 138 m_2 m_4 - 
          14 m_6) U^2 V \nonumber\\& &+
       e^{4 \gamma} (36 m_2^3 + 149 m_3^2 + 282 m_2 m_4 - 
          14 m_6) U V^2-
       e^{6 \gamma} (25 m_3^2 + 46 m_2 m_4 \nonumber\\&& +14 m_6) V^3) - 
    32 m_1^4 m_2 U (U + 
       e^{2 \gamma} V) (10149 U^4 + 
       7633 e^{2 \gamma} U^3 V + 
       799 e^{4 \gamma} U^2 V^2\nonumber\\&  &-
       25 e^{6 \gamma} U V^3 + 
       20 e^{8 \gamma} V^4) + 
    32 m_1^6 U (8721 U^5 + 
       12070 e^{2 \gamma} U^4 V + 
       4722 e^{4 \gamma} U^3 V^2\nonumber\\& & +
       508 e^{6 \gamma} U^2 V^3 + 
       37 e^{8 \gamma} U V^4 + 
       6 e^{10 \gamma} V^5))+\mathcal{O}[\lambda]^7,
\end{eqnarray}
       and
\begin{eqnarray}\label{TrTrg}
    {{T_\mu}}^\mu{T_\nu}^\nu &=&
      16 e^{-4 \gamma}
  \lambda^2   m_1^2(U + e^{2 \gamma} V)^4\nonumber\\& +& 
 64 e^{-5 \gamma}
    \lambda^3 m_1(U + 
    e^{2 \gamma} V)^4 (-4 m_1^2 U + 
    m_2 (U + e^{2 \gamma} V))\nonumber\\&  +&32 e^{-6 \gamma}\lambda^4 (U + 
    e^{2 \gamma}
      V)^4 (-52 m_1^2 m_2 U (U+ 
       e^{2 \gamma} V) + 
    2 m_2^2 (U + e^{2 \gamma} V)^2 \nonumber\\&&+ 
    3 m_1 m_3 (U + e^{2 \gamma} V)^2 + 
    8 m_1^4 U(13 U + 3 e^{2 \gamma}V)) \nonumber\\&+ &
 64 e^{-7 \gamma}\lambda^5 (U + 
    e^{2 \gamma}
      V)^4 (-54 m_1 m_2^2 U (U + 
       e^{2 \gamma} V)^2 - 
    44 m_1^2 m_3 U (U + e^{2 \gamma} V)^2\nonumber\\&  &+
    3 m_2 m_3 (U + e^{2 \gamma} V)^3 + 
    2 m_1 m_4 (U + e^{2 \gamma} V)^3 + 
    96 m_1^3 m_2 U (U + e^{2 \gamma} V) (5 U + 
       e^{2 \gamma} V)\nonumber\\& & -
    32 m_1^5 U(20 U^2 + 11 e^{2 \gamma} U V + 
       e^{4 \gamma} V^2)) \nonumber\\& +& 
 16 e^{-8 \gamma}\lambda^6 (U + 
    e^{2 \gamma}
      V)^4 (208 m_1^3 m_3 U (U + 
       e^{2 \gamma} V)^2 (17 U + 
       3 e^{2 \gamma} V) \nonumber\\&&- 
    608 m_1^4 m_2 U (U + 
       e^{2 \gamma} V) (51 U^2 + 
       25 e^{2 \gamma} U V + 
       2 e^{4 \gamma} V^2) \nonumber\\&&+ 
    32 m_1^6 U (969 U^3 + 
       884 e^{2 \gamma} U^2 V + 
       205 e^{4 \gamma} U V^2 + 
       10 e^{6 \gamma} V^3) \nonumber\\&&- (U + 
       e^{2 \gamma} V)^3 (144 m_2^3 U - 
       9 m_3^2 (U+ e^{2 \gamma} V) - 
       16 m_2 m_4 (U + e^{2 \gamma} V)) \nonumber\\& &-
    2 m_1 (U + e^{2 \gamma} V)^3 (356 m_2 m_3 U - 
       5 m_5 (U + e^{2 \gamma} V))\nonumber\\&  &+
    12 m_1^2 U (U + 
       e^{2 \gamma}
         V)^2 (-22 m_4 (U + e^{2 \gamma} V) + 
       31 m_2^2 (17 U + 
          3 e^{2 \gamma} V)))+[\mathcal{O}]^7,
\end{eqnarray}
\end{appendix}



\bibliographystyle{JHEP}
\bibliography{refs}

\providecommand{\href}[2]{#2}\begingroup\raggedright\begin{thebibliography}{10}

\bibitem{Born:1934gh}
M.~Born and L.~Infeld, {\it {Foundations of the new field theory}},  {\em Proc. Roy. Soc. Lond. A} {\bf 144} (1934), no.~852 425--451.

\bibitem{Born:1933lls}
M.~Born and L.~Infeld, {\it {Electromagnetic mass}},  {\em Nature} {\bf 132} (1933), no.~3347 970.1.

\bibitem{Fradkin:1985qd}
E.~S. Fradkin and A.~A. Tseytlin, {\it {Nonlinear Electrodynamics from Quantized Strings}},  {\em Phys. Lett. B} {\bf 163} (1985) 123--130.

\bibitem{Russo:2024llm}
J.~G. Russo and P.~K. Townsend, {\it {Causal self-dual electrodynamics}},  {\em Phys. Rev. D} {\bf 109} (2024), no.~10 105023, [\href{http://arxiv.org/abs/2401.06707}{{\tt arXiv:2401.06707}}].

\bibitem{Russo:2024xnh}
J.~G. Russo and P.~K. Townsend, {\it {Causality and energy conditions in nonlinear electrodynamics}},  {\em JHEP} {\bf 06} (2024) 191, [\href{http://arxiv.org/abs/2404.09994}{{\tt arXiv:2404.09994}}].

\bibitem{Russo:2024ptw}
J.~G. Russo and P.~K. Townsend, {\it {Dualities of self-dual nonlinear electrodynamics}},  {\em JHEP} {\bf 09} (2024) 107, [\href{http://arxiv.org/abs/2407.02577}{{\tt arXiv:2407.02577}}].

\bibitem{Bialynicki-Birula:1992rcm}
I.~Bialynicki-Birula, {\it {Field theory of photon dust}},  {\em Acta Phys. Polon. B} {\bf 23} (1992) 553--559.

\bibitem{Gaillard:1981rj}
M.~K. Gaillard and B.~Zumino, {\it {Duality Rotations for Interacting Fields}},  {\em Nucl. Phys. B} {\bf 193} (1981) 221--244.

\bibitem{Gaillard:1997rt}
M.~K. Gaillard and B.~Zumino, {\it {Nonlinear electromagnetic selfduality and Legendre transformations}},  in {\em {A Newton Institute Euroconference on Duality and Supersymmetric Theories}}, pp.~33--48, 12, 1997.
\newblock \href{http://arxiv.org/abs/hep-th/9712103}{{\tt hep-th/9712103}}.

\bibitem{Gibbons:1995ap}
G.~W. Gibbons and D.~A. Rasheed, {\it {Sl(2,R) invariance of nonlinear electrodynamics coupled to an axion and a dilaton}},  {\em Phys. Lett. B} {\bf 365} (1996) 46--50, [\href{http://arxiv.org/abs/hep-th/9509141}{{\tt hep-th/9509141}}].

\bibitem{Gibbons:1995cv}
G.~W. Gibbons and D.~A. Rasheed, {\it {Electric - magnetic duality rotations in nonlinear electrodynamics}},  {\em Nucl. Phys. B} {\bf 454} (1995) 185--206, [\href{http://arxiv.org/abs/hep-th/9506035}{{\tt hep-th/9506035}}].

\bibitem{Avetisyan:2021heg}
Z.~Avetisyan, O.~Evnin, and K.~Mkrtchyan, {\it {Democratic Lagrangians for Nonlinear Electrodynamics}},  {\em Phys. Rev. Lett.} {\bf 127} (2021), no.~27 271601, [\href{http://arxiv.org/abs/2108.01103}{{\tt arXiv:2108.01103}}].

\bibitem{Kuzenko:2026kvr}
S.~M. Kuzenko, {\it {On nonlinear self-duality in $4p$ dimensions}},  \href{http://arxiv.org/abs/2601.13022}{{\tt arXiv:2601.13022}}.

\bibitem{10.1115/1.3630089}
Courant and Hilbert, {\it Methods of mathematical physics, vol. ii. partial differential equations},  {\em Journal of Applied Mechanics} {\bf 30} (03, 1963) 158--158.

\bibitem{Ivanov:2002ab}
E.~A. Ivanov and B.~M. Zupnik, {\it {New representation for Lagrangians of selfdual nonlinear electrodynamics}},  in {\em {4th International Workshop on Supersymmetry and Quantum Symmetries}: {16th Max Born Symposium}}, pp.~235--250, 2002.
\newblock \href{http://arxiv.org/abs/hep-th/0202203}{{\tt hep-th/0202203}}.

\bibitem{Ivanov:2003uj}
E.~A. Ivanov and B.~M. Zupnik, {\it {New approach to nonlinear electrodynamics: Dualities as symmetries of interaction}},  {\em Phys. Atom. Nucl.} {\bf 67} (2004) 2188--2199, [\href{http://arxiv.org/abs/hep-th/0303192}{{\tt hep-th/0303192}}].

\bibitem{Mkrtchyan:2022ulc}
K.~Mkrtchyan and M.~Svazas, {\it {Solutions in Nonlinear Electrodynamics and their double copy regular black holes}},  {\em JHEP} {\bf 09} (2022) 012, [\href{http://arxiv.org/abs/2205.14187}{{\tt arXiv:2205.14187}}].

\bibitem{Babaei-Aghbolagh:2024uqp}
H.~Babaei-Aghbolagh, S.~He, and H.~Ouyang, {\it {Generalized $ T\overline{T} $-like deformations in duality-invariant nonlinear electrodynamic theories}},  {\em JHEP} {\bf 09} (2024) 137, [\href{http://arxiv.org/abs/2407.03698}{{\tt arXiv:2407.03698}}].

\bibitem{Babaei-Aghbolagh:2025cni}
H.~Babaei-Aghbolagh, K.~Babaei~Velni, S.~He, and Z.~Pezhman, {\it {Root-$ T\overline{T} $ deformations on causal self-dual electrodynamic theories}},  {\em JHEP} {\bf 07} (2025) 227, [\href{http://arxiv.org/abs/2504.10361}{{\tt arXiv:2504.10361}}].

\bibitem{Russo:2025fuc}
J.~G. Russo and P.~K. Townsend, {\it {Simplified self-dual electrodynamics}},  {\em JHEP} {\bf 10} (2025) 120, [\href{http://arxiv.org/abs/2505.08869}{{\tt arXiv:2505.08869}}].

\bibitem{Smirnov:2016lqw}
F.~A. Smirnov and A.~B. Zamolodchikov, {\it {On space of integrable quantum field theories}},  {\em Nucl. Phys. B} {\bf 915} (2017) 363--383, [\href{http://arxiv.org/abs/1608.05499}{{\tt arXiv:1608.05499}}].

\bibitem{Cavaglia:2016oda}
A.~Cavagli{\`a}, S.~Negro, I.~M. Sz{\'e}cs{\'e}nyi, and R.~Tateo, {\it {$T \bar{T}$-deformed 2D Quantum Field Theories}},  {\em JHEP} {\bf 10} (2016) 112, [\href{http://arxiv.org/abs/1608.05534}{{\tt arXiv:1608.05534}}].

\bibitem{Conti:2018jho}
R.~Conti, L.~Iannella, S.~Negro, and R.~Tateo, {\it {Generalised Born-Infeld models, Lax operators and the $ \mathrm{T}\overline{\mathrm{T}} $ perturbation}},  {\em JHEP} {\bf 11} (2018) 007, [\href{http://arxiv.org/abs/1806.11515}{{\tt arXiv:1806.11515}}].

\bibitem{Babaei-Aghbolagh:2020kjg}
H.~Babaei-Aghbolagh, K.~Babaei~Velni, D.~M. Yekta, and H.~Mohammadzadeh, {\it {$ T\overline{T} $-like flows in non-linear electrodynamic theories and S-duality}},  {\em JHEP} {\bf 04} (2021) 187, [\href{http://arxiv.org/abs/2012.13636}{{\tt arXiv:2012.13636}}].

\bibitem{Babaei-Aghbolagh:2022uij}
H.~Babaei-Aghbolagh, K.~B. Velni, D.~M. Yekta, and H.~Mohammadzadeh, {\it {Emergence of non-linear electrodynamic theories from TT{\textasciimacron}-like deformations}},  {\em Phys. Lett. B} {\bf 829} (2022) 137079, [\href{http://arxiv.org/abs/2202.11156}{{\tt arXiv:2202.11156}}].

\bibitem{Babaei-Aghbolagh:2022itg}
H.~Babaei-Aghbolagh, K.~Babaei~Velni, D.~M. Yekta, and H.~Mohammadzadeh, {\it {Manifestly SL(2, R) Duality-Symmetric Forms in ModMax Theory}},  {\em JHEP} {\bf 12} (2022) 147, [\href{http://arxiv.org/abs/2210.13196}{{\tt arXiv:2210.13196}}].

\bibitem{Ferko:2023wyi}
C.~Ferko, S.~M. Kuzenko, L.~Smith, and G.~Tartaglino-Mazzucchelli, {\it {Duality-invariant nonlinear electrodynamics and stress tensor flows}},  {\em Phys. Rev. D} {\bf 108} (2023), no.~10 106021, [\href{http://arxiv.org/abs/2309.04253}{{\tt arXiv:2309.04253}}].

\bibitem{Ferko:2024zth}
C.~Ferko, S.~M. Kuzenko, K.~Lechner, D.~P. Sorokin, and G.~Tartaglino-Mazzucchelli, {\it {Interacting chiral form field theories and $ T\overline{T} $-like flows in six and higher dimensions}},  {\em JHEP} {\bf 05} (2024) 320, [\href{http://arxiv.org/abs/2402.06947}{{\tt arXiv:2402.06947}}].

\bibitem{Hutomo:2025dfx}
J.~Hutomo, K.~Lechner, and D.~P. Sorokin, {\it {On non-linear chiral 4-form theories in D=10}},  \href{http://arxiv.org/abs/2509.14351}{{\tt arXiv:2509.14351}}.

\bibitem{Babaei-Aghbolagh:2025hlm}
H.~Babaei-Aghbolagh, B.~Chen, and S.~He, {\it {Integrable Sigma Models and Universal Root $T\bar{T}$ Deformation via Courant-Hilbert Approach}},  \href{http://arxiv.org/abs/2509.17075}{{\tt arXiv:2509.17075}}.

\bibitem{Babaei-Aghbolagh:2025uoz}
H.~Babaei-Aghbolagh, B.~Chen, and S.~He, {\it {Root-$T\bar{T}$ Flows Unify 4D Duality-Invariant Electrodynamics and 2D Integrable Sigma Models}},  \href{http://arxiv.org/abs/2507.22808}{{\tt arXiv:2507.22808}}.

\bibitem{Bandos:2020jsw}
I.~Bandos, K.~Lechner, D.~Sorokin, and P.~K. Townsend, {\it {A non-linear duality-invariant conformal extension of Maxwell's equations}},  {\em Phys. Rev. D} {\bf 102} (2020) 121703, [\href{http://arxiv.org/abs/2007.09092}{{\tt arXiv:2007.09092}}].

\bibitem{Ferko:2022iru}
C.~Ferko, L.~Smith, and G.~Tartaglino-Mazzucchelli, {\it {On Current-Squared Flows and ModMax Theories}},  {\em SciPost Phys.} {\bf 13} (2022), no.~2 012, [\href{http://arxiv.org/abs/2203.01085}{{\tt arXiv:2203.01085}}].

\bibitem{Babaei-Aghbolagh:2025lko}
H.~Babaei-Aghbolagh, S.~He, and H.~Ouyang, {\it {Generalized TT{\textasciimacron}-like flows for scalar theories in two dimensions}},  {\em Phys. Rev. D} {\bf 112} (2025), no.~6 066005, [\href{http://arxiv.org/abs/2501.14583}{{\tt arXiv:2501.14583}}].

\bibitem{Kuzenko:2025gvn}
S.~M. Kuzenko and J.~Ruhl, {\it {Generalisations of the Russo-Townsend formulation}},  \href{http://arxiv.org/abs/2511.20051}{{\tt arXiv:2511.20051}}.

\bibitem{Babaei-Aghbolagh:2022leo}
H.~Babaei-Aghbolagh, K.~Babaei~Velni, D.~Mahdavian~Yekta, and H.~Mohammadzadeh, {\it {Marginal TT{\textasciimacron}-like deformation and modified Maxwell theories in two dimensions}},  {\em Phys. Rev. D} {\bf 106} (2022), no.~8 086022, [\href{http://arxiv.org/abs/2206.12677}{{\tt arXiv:2206.12677}}].

\bibitem{Ferko:2022cix}
C.~Ferko, A.~Sfondrini, L.~Smith, and G.~Tartaglino-Mazzucchelli, {\it {Root-$T \bar T$ Deformations in Two-Dimensional Quantum Field Theories}},  {\em Phys. Rev. Lett.} {\bf 129} (2022), no.~20 201604, [\href{http://arxiv.org/abs/2206.10515}{{\tt arXiv:2206.10515}}].

\bibitem{Borsato:2022tmu}
R.~Borsato, C.~Ferko, and A.~Sfondrini, {\it {Classical integrability of root-TT{\textasciimacron} flows}},  {\em Phys. Rev. D} {\bf 107} (2023), no.~8 086011, [\href{http://arxiv.org/abs/2209.14274}{{\tt arXiv:2209.14274}}].

\bibitem{Ferko:2025bhv}
C.~Ferko, M.~Galli, Z.~Huang, and G.~Tartaglino-Mazzucchelli, {\it {Soliton Surfaces and the Geometry of Integrable Deformations of the $\mathbb{CP}^{N-1}$ Model}},  \href{http://arxiv.org/abs/2509.05081}{{\tt arXiv:2509.05081}}.

\bibitem{Baglioni:2025tsc}
N.~Baglioni, D.~Bielli, M.~Galli, and G.~Tartaglino-Mazzucchelli, {\it {Relating auxiliary field formulations of $4d$ duality-invariant and $2d$ integrable field theories}},  \href{http://arxiv.org/abs/2512.21982}{{\tt arXiv:2512.21982}}.

\bibitem{Sakamoto:2025hwi}
J.-i. Sakamoto, R.~Tateo, and M.~Yamazaki, {\it {$T\overline{T}$ and root-$T\overline{T}$ deformations in four-dimensional Chern-Simons theory}},  {\em JHEP} {\bf 01} (2026) 084, [\href{http://arxiv.org/abs/2509.12303}{{\tt arXiv:2509.12303}}].

\bibitem{Fukushima:2025tlj}
O.~Fukushima, T.~Matsumoto, and K.~Yoshida, {\it {The Courant-Hilbert construction in 4D Chern-Simons theory}},  \href{http://arxiv.org/abs/2509.22080}{{\tt arXiv:2509.22080}}.

\end{thebibliography}\endgroup

\end{document}